\documentclass[article, shortnames]{jss}
\usepackage{thumbpdf,lmodern,bm,amsmath,bbm,algorithm,amsfonts}
\usepackage{algpseudocode}

\usepackage{framed}

\newcommand{\fct}[1]{\code{#1()}}

\makeatletter
\newcommand{\algmargin}{\the\ALG@thistlm}   
\makeatother
\algnewcommand{\parState}[1]{\State%
    \parbox[t]{\dimexpr\linewidth-\algmargin}{\strut\hangindent=\algorithmicindent \hangafter=1 #1\strut}}

\author{Chuji Luo\\University of Florida
   \And Michael J. Daniels\\University of Florida}
\Plainauthor{Chuji Luo, Michael J. Daniels}

\title{\pkg{BNPqte}: A Bayesian Nonparametric Approach to Causal Inference on Quantiles in \proglang{R}}
\Plaintitle{BNPqte: A Bayesian Nonparametric Approach to Causal Inference on Quantiles in R}
\Shorttitle{\pkg{BNPqte}: A Bayesian Nonparametric Approach to Causal Inference on Quantiles in \proglang{R}}

\Abstract{
  In this article, we introduce the \pkg{BNPqte} \proglang{R} package which implements the Bayesian nonparametric approach of \cite{xu2018bayesian} for estimating quantile treatment effects in observational studies. This approach provides flexible modeling of the distributions of potential outcomes, so it is capable of capturing a variety of underlying relationships among the outcomes, treatments and confounders and estimating multiple quantile treatment effects simultaneously. Specifically, this approach uses a Bayesian additive regression trees (BART) model to estimate the propensity score and a Dirichlet process mixture (DPM) of multivariate normals model to estimate the conditional distribution of the potential outcome given the estimated propensity score. The \pkg{BNPqte} \proglang{R} package provides a fast implementation for this approach by designing efficient \proglang{R} functions for the DPM of multivariate normals model in joint and conditional density estimation. These \proglang{R} functions largely improve the efficiency of the DPM model in density estimation, compared to the popular \pkg{DPpackage}. BART-related \proglang{R} functions in the \pkg{BNPqte} \proglang{R} package are inherited from the \pkg{BART} \proglang{R} package with two modifications on variable importance and split probability. To maximize computational efficiency, the actual sampling and computation for each model are carried out in \proglang{C++} code. The \pkg{Armadillo} \proglang{C++} library is also used for fast linear algebra calculations.
}

\Plainkeywords{Bayesian additive regression trees (BART), Bayesian nonparametric (BNP), Dirichlet process mixture (DPM), Bayesian bootstrap, propensity score (PS), quantile causal effect (QTE), R, RcppArmadillo, OpenMP}

\begin{document}


\section[Introduction]{Introduction} \label{sec:intro}

Estimating causal effects is the objective of causal inference. Commonly used average treatment effects are often of less interest when the distribution of the outcome is skewed or multi-modal, and quantile treatment effects become more sensible and desirable. Under the potential outcome framework (\cite{rubin1978bayesian}, \cite{splawa1990application}) and the binary treatment setting, the $p^{\text{th}}$ quantile treatment effect (QTE) is defined as the difference between the $p^{\text{th}}$ population quantiles of the treated and untreated potential outcome distributions, i.e.,
\begin{equation} \label{eq:0}
F^{-1}_1(p) \, - \, F^{-1}_0(p), \qquad p \in (0,1),
\end{equation}
where $F^{-1}_t(p)$, $t \in \{0, 1\}$, is the $p^{\text{th}}$ quantile for the cumulative distribution function (CDF) $F_t(y)$ of the potential outcome $Y(t)$ with $Y(1)$ and $Y(0)$ denoting the treated and untreated potential outcomes, respectively. There has been extensive work from both the frequentist and Bayesian perspectives (\cite{chernozhukov2005iv}, \cite{zhang2012causal}, \cite{hoshino2013semiparametric}, \cite{linn2014interactive}, \cite{karabatsos2015bayesian}, \cite{diaz2017efficient}) on estimating QTEs for observational data. They typically develop their own identifying assumptions and many require different sets of efficient estimating equations for different quantiles.

\cite{xu2018bayesian} propose a fully Bayesian nonparametric (BNP) approach to estimate QTEs under a set of mild identifying assumptions. Specifically, they model the distributions of potential outcomes by first estimating the propensity score (PS) using a probit Bayesian additive regression trees (BART) model (\cite{chipman2010bart}), then modeling the conditional distribution of the potential outcome given a BART posterior sample of the PS in each treatment group separately using a Dirichlet process mixture (DPM) of multivariate normals model (\cite{muller1996bayesian}) and finally marginalizing the estimated conditional distribution over the population distribution of the confounders using Bayesian bootstrap (\cite{rubin1981bayesian}). The recovery of the distributions of potential outcomes enables the inference on any functionals of the distributions, including estimating multiple QTEs simultaneously. Additionally, the BNP methods provide a flexible approach to capture a wide range of relationships among the outcomes, treatments and confounders. Furthermore, modeling the distribution of the potential outcome conditional on the PS rather than on the confounders avoids potentially complex high-dimensional modeling and the two-stage approach of modeling the PS and the conditional distribution of the potential outcome on the PS separately averts the model feedback issue discussed in \cite{zigler2013model}. Finally, the use of posterior samples of the PS rather than a point estimator of the PS propagates the uncertainty in the estimated PS. For convenience, throughout this article, we refer to this BNP approach as the BART-DPM approach.

On the Comprehensive \proglang{R} Archive Network (CRAN), there are many \proglang{R} packages implementing either the frequentist or Bayesian methods for estimating the average treatment effects, including (but not limited to) \pkg{ipw} (\cite{van2011ipw}), \pkg{pcalg} (\cite{kalisch2012causal}), \pkg{CausalImpact} (\cite{brodersen2015inferring}) and  \pkg{causaleffect} (\cite{tikka2018identifying}). However, only two \proglang{R} packages, \pkg{qte} \citep{Callaway2019} and \pkg{Counterfactual} \citep{Chen2020}, to our knowledge, are designed to estimate QTEs and the methods implemented are all frequentist. We bridge the gap by implementing the BART-DPM approach, a fully Bayesian nonparametric approach, in the \pkg{BNPqte} \proglang{R} \citep{R} package.

The BART-DPM approach includes the application of BART and DPM models. Currently, there are four CRAN packages available for the implementation of BART: \pkg{bartMachine} (\cite{kapelner2016}), \pkg{BayesTree} \citep{Chipman2016}, \pkg{dbarts} \citep{Dorie2020} and \pkg{BART} (\cite{sparapani2021nonparametric}). The \pkg{bartMachine} \proglang{R} package is written in \proglang{JAVA}, but the \proglang{R} to \proglang{JAVA} interface can be challenging to deal with as \proglang{R} is mainly written in \proglang{C++} and \proglang{Fortran}. The \pkg{BayesTree} and \pkg{dbarts} (a clone of \pkg{BayesTree} interface) \proglang{R} packages are written in \proglang{C++}, but there are no future improvements planned for \pkg{BayesTree} so the packages following it, such as \pkg{dbarts}, are also strongly affected. The \pkg{BART} \proglang{R} package also chooses \proglang{C++} as the main programming language, because both of \proglang{R} and \proglang{C++} are object-oriented languages and seamless \proglang{R} and \proglang{C++} integration can be provided by the \pkg{Rcpp} (\cite{Dirk2011}) packages. Moreover, \pkg{BART} is actively maintained. As such, in the \pkg{BNPqte} package, we adopt \pkg{BART} (specifically, its functions \fct{wbart}, \fct{pbart} and \fct{lbart}) for the implementation of BART models. Slight modifications are applied to these functions to adjust for the data with mixed-type covariates and also to take into account the optimal posterior contraction for the BART prior, the latter of which is discussed in \cite{rovckova2019theory}.

Exact posterior distributions of DPM models are analytically intractable, so the implementations of DPM models heavily rely on the MCMC chain Monte Carlo (MCMC) samplers for the posterior distributions of DPM models. There are several CRAN packages implementing DPM models with different sampling algorithms and different purposes. \pkg{DPpackage} (\cite{jara2011dppackage}), an archived CRAN package, implements Algorithm 4 and Algorithm 8 of \cite{neal2000markov}, the former of which is essentially the "no-gaps" algorithm of \cite{maceachern1998estimating}. The \pkg{dirichletprocess} \citep{Ross2020} \proglang{R} package also employs these two samplers but with a different purpose. Both the two samplers are based on the Pólya urn scheme and explore the marginal parameter space where some parameters are integrated out. Although these two samplers mitigate the slow mixing issue of the algorithm proposed in \cite{escobar1995bayesian}, they do not have the direct access to the full posterior distribution and therefore require complicated algorithms to obtain it. \cite{ishwaran2001gibbs} consider a truncation approximation based on a mixture of Dirichlet process and suggest a blocked Gibbs sampling method that directly samples from the full posterior distribution of the truncated DPM model. This sampler leads to a rapidly mixing Markov chain and also permits direct inference on the full posterior distribution of the truncated DPM model. The \pkg{PReMiuM} (\cite{liverani2015premium}) \proglang{R} package implements this sampler, but the application of DPM models in \pkg{PReMiuM} is dedicated to the profile regression, which cannot be directly applied to other tasks such as density estimation which is relevant to the work here. Out of the consideration of the computational efficiency and the convergence speed of Markov chains, we implement the blocked Gibbs sampler of \cite{ishwaran2001gibbs} and Algorithm 8  (with $m=1$) of \cite{neal2000markov} in the \pkg{BNPqte} package for the DPM of multivariate normals model with application to nonparametric density estimation.

This article is organized as follows. In Section~\ref{sec:bart}, we briefly review BART models and illustrate the modifications that we make for the \proglang{R} functions inherited from the \pkg{BART} package with a simulated example. In Section~\ref{sec:dpmm}, we describe the DPM of multivariate normals model and two MCMC samplers implemented in the \pkg{BNPqte} package. We also demonstrate the usage of the DPM-related \proglang{R} functions in density estimation with simulated examples in Section~\ref{sec:dpmm}. In Section~\ref{sec:bnp}, we introduce the BART-DPM approach for estimating QTEs, which is the focus of the \pkg{BNPqte} package, and present the usage of the \pkg{BNPqte} package in QTE estimation with a simulated example from \cite{xu2018bayesian}. We investigate the convergence of Markov chains from DPM models in Section~\ref{sec:diagnostics} and finally conclude this article with a summary in Section~\ref{sec:summary}. More computational details are provided in Appendix.


\section{BNP Regression with BART Models} \label{sec:bart}

BART, built on the Bayesian classification and regression tree algorithm (\cite{chipman1998bayesian}), is a Bayesian approach to nonparametric function estimation and inference using a sum of trees. Since \cite{chipman2010bart}, BART has been widely studied, including extensions to generalized regression and high-dimensional data, applications to variable selection and causal inference, theoretical studies and more. In this section, we first review the BART models for continuous and binary responses, BART-based variable importance and recent theoretical results, which are the implemented specification of BART in the \pkg{BNPqte} package, and then illustrate the usage of BART-related \proglang{R} functions with a simulated example.


\subsection{BART Models Specification}

Consider the regression problem of making inference about an unknown function $f_0$ that predicts a continuous response $Y_i$ using a $p$-dimensional vector of predictors $\boldsymbol{x}_i = (x_{i1}, \cdots, x_{ip})^{\top}$ ($1\leq i\leq n$),
\begin{equation} \label{eq:1}
Y_i \, =  \, f_0(\boldsymbol{x}_i) + \epsilon_i, \qquad \epsilon_i \, \stackrel{\text{iid}}{\sim} \, \mathrm{Normal}(0, \sigma^2).
\end{equation}
BART models $f_0$ by a sum of $M$ Bayesian regression trees and can be expressed as
\begin{align} \label{eq:2}
Y_i \, &= \, f_{\text{BART}}(\boldsymbol{x}_i) + \epsilon_i, \qquad \epsilon_i \, \stackrel{\text{iid}}{\sim} \, \mathrm{Normal}(0, \sigma^2), \nonumber \\
f(\boldsymbol{x}_i) \, &= \, \sum\limits_{m=1}^M g(\boldsymbol{x}_i; \mathcal{T}_m, \boldsymbol{\mu}_m),
\end{align}
where $g(\boldsymbol{x}_i; \mathcal{T}_m, \boldsymbol{\mu}_m)$ is the output of $\boldsymbol{x}_i$ from a single regression tree. Each $g(\boldsymbol{x}_i; \mathcal{T}_m, \boldsymbol{\mu}_m)$ is determined by the binary tree structure $\mathcal{T}_m$ that consists of a set of splitting rules and a set of terminal nodes and the vector of parameters $\boldsymbol{\mu}_m = (\mu_{m,1}, \cdots \mu_{m,b_m})$ associated with each of the $b_m$ terminal nodes of $\mathcal{T}_m$, such that $g(\boldsymbol{x}_i; \mathcal{T}_m, \boldsymbol{\mu}_m) = \mu_{m,l}$ if $\boldsymbol{x}_i$ is associated with the $l^\text{th}$ terminal node of the tree $m$.

The prior of BART is specified for three components: (1) the $M$ independent trees structures $\{\mathcal{T}_m\}_{m=1}^M$, (2) the parameters $\{\boldsymbol{\mu}_m\}_{m=1}^M$ associated with the terminal nodes given $\{\mathcal{T}_m\}_{m=1}^M$, and (3) the error variance $\sigma^2$ that is independent with the former two. A regularization prior is placed on the ensemble structure $\{\mathcal{T}_m\}_{m=1}^M$. It consists of a Bernoulli distribution with probability
\begin{equation} \label{eq:3}
p_{\text{split}}(d) \, = \, \frac{\gamma}{(1+d)^{\beta}}, \quad\quad \gamma \in (0,1),~\beta \in (0, \infty),
\end{equation}
for splitting a node at depth $d$ ($d \in \{0, 1, \cdots\}$) into two child nodes and two discrete uniform distributions for selecting a split variable and a split value given the selected split variable. This regularization prior keeps an individual tree from being too influential, thereby improving the overall fit and avoiding potential overfitting problems. Data-informed conjugate normal and inverse gamma priors are used for $\mu_{m,l}$'s and $\sigma^2$, respectively.

The posterior distribution is sampled through a Metropolis-within-Gibbs sampler (\cite{hastings1970monte}, \cite{geman1984stochastic}) which uses Bayesian backfitting (\cite{hastie2000bayesian}) to update each tree iteratively. Estimation of $f_0(\boldsymbol{x}_i)$ is the average of BART posterior samples of $f_{\text{BART}}(\boldsymbol{x}_i)$ after a burn-in period. More details of the BART prior, posterior and estimation computation can be found in the original BART paper (\cite{chipman2010bart}).

The continuous BART model (\ref{eq:2}) has been extended to a variety of regression models for non-continuous responses, including probit BART (\cite{chipman2010bart}) and logit BART (\cite{sparapani2021nonparametric}) for binary responses, which along with continuous BART, are the three BART models implemented in the \pkg{BNPqte} package.

Probit BART and logit BART specify
\begin{equation} \label{eq:4}
P(Y_i = 1 \, | \, \boldsymbol{x}_i) \, =  \, H[f_{\text{BART}}(\boldsymbol{x}_i)],
\end{equation}
where $f_{\text{BART}}(\boldsymbol{x}_i)$ is the sum-of-trees function in (\ref{eq:2}) and $H$ is the link function with the probit link for probit BART and the logistic link for logit BART. Both of the models assign the same prior to the ensemble structure and the parameters of the terminal nodes, i.e., $\{\mathcal{T}_m, \boldsymbol{\mu}_m\}_{m=1}^M$, as continuous BART, but $\sigma^2$ is fixed due to identifiability.

Probit BART uses the data augmentation approach of \cite{albert1993bayesian} to adapt the Bayesian backfitting sampler of continuous BART. Specifically, a latent variable $Z_i$ such that $Y_i=\mathbbm{1}(Z_i>0)$ is first introduced for each response variable $Y_i$ and can be imputed by sampling from the full conditional distribution of $Z_i$ given $Y_i$ and other parameters at each iteration of the MCMC algorithm. The full conditional distribution essentially is a truncated normal distribution. The imputed $Z_i$'s are then modeled by the continuous BART model with $\sigma^2=1$, so the MCMC algorithm can be completed by performing the Bayesian backfitting algorithm of the continuous BART model on the imputed $Z_i$'s.

Logit BART also introduces latent $Z_i$'s but assumes them to follow a logistic distribution which has a heavier tail than a normal distribution, thereby improving the estimation for extreme $P(Y_i = 1 \, | \, \boldsymbol{x}_i)$. The latent $Z_i$'s are sampled using the technique of \cite{gramacy2012simulation} and conditional on the imputed $Z_i$'s, the continuous BART model with given heteroskedastic variance $\sigma_i^2$'s are fitted on $Z_i$'s, where $\sigma_i^2$'s are obtained by the method of \cite{robert1995simulation}.
Details of the computation for probit and logit BART can be found in \cite{sparapani2021nonparametric}.
Though both probit BART and logit BART can be applied to the binary regression problem, it is important to recognize that probit BART is more computationally efficient than logit BART.


\subsection{BART-Based Variable Importance} \label{sec:vi}

BART has also been applied to variable selection (\cite{bleich2014variable}, \cite{linero2018bayesian}, \cite{liu2018variable}, \cite{luo2021}). The \pkg{BNPqte} package provides five types of BART variable importance measures which can be used for variable selection; three of them are inherited from the \pkg{BART} package.

The three variable importance measures provided by the \pkg{BART} package are: variable inclusion proportion (VIP), marginal posterior variable inclusion probability (PVIP) and posterior split probability (PSP). VIP, proposed by \cite{chipman2010bart}, describes the average use per splitting rule for each predictor. The latter two variable importance measures are the products of the DART model (\cite{linero2018bayesian}) which modifies BART by replacing the discrete uniform distribution for selecting a split variable with a categorical distribution of which the event probabilities follow a Dirichlet distribution. PVIP estimates the average times that a predictor is included in an ensemble at least once and PSP suggests the probability that a predictor is selected as a split variable for a splitting rule. VIP and PSP are used as variable importance with the idea that predictors are important if they are used for many splitting rules. However, this idea fails in the presence of different types of predictors. For example, an important binary predictor can only be used once at most in each path of a tree because it only has two unique values. Also, the use of PVIP in variable selection often requires strong assumptions and the performance becomes unstable when predictors are correlated.

\pkg{BNPqte} also implements two variable importance measures proposed in \cite{luo2021}, which are designed for data with mixed-type predictors. The first one is the within-type variable inclusion proportion (within-type VIP) which describes the average use per splitting rule for a predictor, out of those splitting rules using the same type of predictors. Within-type VIP shows the relative importance of predictors within each type and improves the permutation-based variable selection approach (\cite{bleich2014variable}). The second variable importance measure is the Metropolis importance (MI) which is defined as the average Metropolis acceptance ratio per splitting rule out of the splitting rules using the predictor. This variable importance helps identify all the relevant predictors for data with general mixed-type predictors. More discussions on these variable importance measures can be found in \cite{luo2021}.


\subsection{Theoretical Properties of BART Priors}

Theoretical exploration of BART has begun in recent years, including \cite{linero2017bayesian}, \cite{rovckova2019theory}, \cite{rovckova2020posterior} and \cite{jeong2020art}. \cite{linero2017bayesian} propose the soft BART prior to adapt to unknown smoothness and sparsity levels and show a near-optimal posterior contraction rate for sparse functions with an unknown smoothness level, or functions with additive structures. \cite{rovckova2020posterior} develop the spike-and-tree prior which wraps the Bayesian CART prior (\cite{denison1998bayesian}) with a spike-and-slab prior for dimension reduction and variable selection, and provide a near-optimal posterior contraction rate also for sparse function but with an unknown smoothness level up to $1$, or functions with additive structures. \cite{jeong2020art} modify the spike-and-tree prior by split-nets which remove the prior dependence on the observed covariates values, and further establish theoretical results for a wide range of functions which can have anisotropic smoothness that possibly varies over the function domain. While these works study variants of the BART prior of \cite{chipman2010bart}, which is referred to as the "exact BART" prior, only \cite{rovckova2019theory} focus on the exact BART prior. They show that it is not clear whether the exact BART prior that uses (\ref{eq:3}) as the split probability has the optimal tail property of the number of terminal nodes and therefore that the optimal posterior contraction rate is not guaranteed. Instead, they show that the optimal tail property can be obtained if (\ref{eq:3}) is replaced by the split probability
\begin{equation} \label{eq:5}
p_{\text{split}}(d) \, = \, \gamma^d,
\end{equation}
for some $\gamma \in [1/n, 1/2)$. As such, the model using the exact BART prior with the slightly modified split probability (\ref{eq:5}) can achieve the optimal posterior contraction (up to a log factor) for $\nu$-H\"{o}lder continuous regression functions with $0<\nu\leq1$,  as shown in in Theorem 7.1 of \cite{rovckova2019theory}. The \pkg{BNPqte} package implements both the two types of split probability, (\ref{eq:3}) and (\ref{eq:5}), in the BART-related \proglang{R} functions.


\subsection{Illustrations}

The BART-related \proglang{R} functions in the \pkg{BNPqte} package are inherited from the \pkg{BART} package: \fct{wbart} for continuous BART, \fct{pbart} for probit BART and \fct{lbart} for logit BART. While the original features of these functions are preserved, two modifications are made. We illustrate the modifications by taking the function \fct{wbart} as an example. The usage of \fct{pbart} and \fct{lbart} is similar because the modifications are made only on trees structures which are shared by the three models.

The first modification is to add the split probability (\ref{eq:5}) as an alternative to the original split probability (\ref{eq:3}). When calling \fct{wbart}, one can specify the split probability to be either (\ref{eq:3}) by setting the argument \code{split.prob = "polynomial"} or (\ref{eq:5}) by setting \code{split.prob = "exponential"}. The character strings \code{"polynomial"} and \code{"exponential"} are named after the decreasing rate of the corresponding split probability with respect to the depth. The default values of the hyper-parameters $\gamma$ and $\beta$ in (\ref{eq:3}) are set the same as \pkg{BART}, i.e., \code{base = 0.95} for $\gamma$ and \code{power = 2} for $\beta$. When \code{split.prob = "exponential"}, the argument \code{power} will not be used and the argument \code{base} is used to represent $\gamma$ in (\ref{eq:5}). In this case, we set \code{base = 0.5} by default. Other arguments of \fct{wbart} in our package are the same as that in the \pkg{BART} package.

The second modification is to add two new variable importance measures, within-type VIP and MI, to these functions, in order to improve variable selection in the presence of mixed-type predictors, as discussed in Section~\ref{sec:vi}. The two new variable importance measures, similar to the original three variable importance measures, can be obtained from the output of \fct{wbart}, as shown below:
\begin{itemize}
  \item \code{vip} is a vector of variable inclusion proportions (VIP) proposed in \cite{chipman2010bart}
  \item \code{within_type_vip} is a vector of within-type VIPs proposed in \cite{luo2021}
  \item \code{mi} is a vector of Metropolis importance (MI) proposed in \cite{luo2021}
  \item \code{pvip} is a vector of marginal posterior variable inclusion probabilities (PVIP) proposed in \cite{linero2018bayesian}; only useful when DART is fit, i.e., \code{sparse = TRUE}
  \item \code{varprob.mean} is a vector of posterior split probabilities (PSP) proposed in \cite{linero2018bayesian}; only useful when DART is fit, i.e., \code{sparse = TRUE}.
\end{itemize}

Now, we demonstrate the two newly added features with a simulated example. We simulate $500$ observations with $5$ binary predictors $x_1, \cdots, x_5$ from $\mathrm{Bernoulli}(0.5)$ and $5$ continuous predictors $x_6,\cdots, x_{10}$ from $\mathrm{Uniform}(0,1)$. The continuous response $Y$ is sampled from $\mathrm{Normal}(f_0(\boldsymbol{x}), 1)$, where
\begin{equation} \label{exp:1}
f_0(\boldsymbol{x}) \, = \, 10\sin (\pi x_1 x_6) + 20 (x_8 - 0.5)^2 + 10 x_2 + 5 x_7.
\end{equation}
The data generation process described above can be done by using the function \fct{MixData} in our package as follows.
\begin{Schunk}
\begin{Sinput}
R> library(BNPqte)
R> set.seed(0)
R> BartData = MixData(n = 500, p = 10, sigma = 1, binary = F)
\end{Sinput}
\end{Schunk}

\begin{figure}[h!]
\centering
\includegraphics{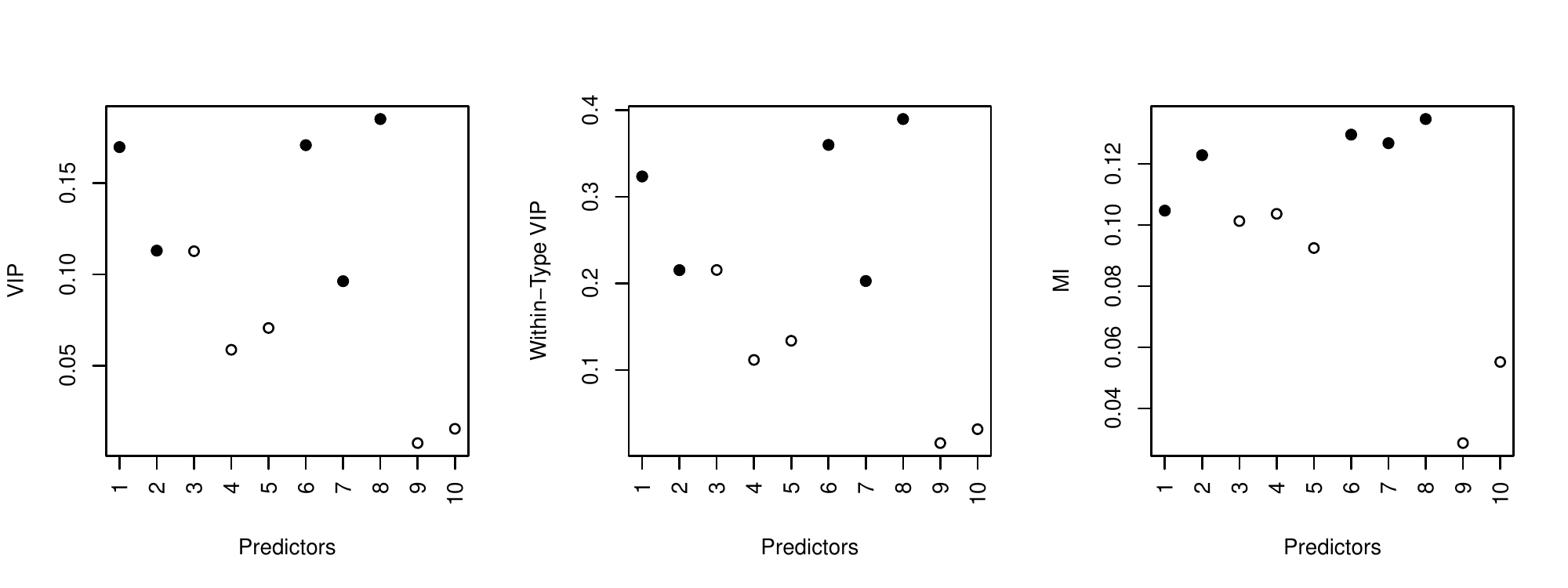}
\caption{\label{fig:bart}BART-based variable importance measures, VIP, within-type VIP and MI, from the BART model using the split probability (\ref{eq:3}) for the example in (\ref{exp:1}). Solid points are for actual relevant predictors and empty points are for irrelevant predictors.}
\end{figure}

We fit two BART models on the data (\code{BartData$Y}, \code{BartData$X}) with split probability (\ref{eq:3}) and (\ref{eq:5}), respectively. Specifically, the BART models are fit with $50$ trees (i.e., \code{ntree = 50}) and $500$ posterior samples are kept (i.e., \code{ndpost = 500}) after burning in $100$ samples (i.e., \code{nskip =100}). Other hyper-parameters are specified as default.
\begin{Schunk}
\begin{Sinput}
R> BartPostP = wbart(BartData$X, BartData$Y, split.prob = "polynomial",
+                    ntree = 50L, ndpost = 500L, nskip = 100L)
R> BartPostE = wbart(BartData$X, BartData$Y, split.prob = "exponential",
+                    ntree = 50L, ndpost = 500L, nskip = 100L)
\end{Sinput}
\end{Schunk}
The results of running \fct{wbart} using split probability (\ref{eq:3}) and (\ref{eq:5}) are stored in the object \code{BartPostP} and \code{BartPostE}, respectively, both of which are of type \code{"wbart"}.  In a comparison with the root mean squared errors (RMSE), the BART model using the split probability (\ref{eq:5}) is slightly preferred, as shown below.
\begin{Schunk}
\begin{Sinput}
R> cat("RMSE of BartPostP: ", 
+      sqrt(mean((BartData$Y - BartPostP$yhat.train.mean)^2)), "\n",
+      "RMSE of BartPostE: ", 
+      sqrt(mean((BartData$Y - BartPostE$yhat.train.mean)^2)), sep = "")
\end{Sinput}
\begin{Soutput}
RMSE of BartPostP: 0.9203228
RMSE of BartPostE: 0.8706075
\end{Soutput}
\end{Schunk}
The three BART-based variable importance measures, MI, VIP and within-type VIP, can be obtained from an object of type \code{"wbart"}. In this example, MI, VIP and within-type VIP of the BART model using split probability (\ref{eq:3}) are stored in \code{BartPostP\$mi}, \code{BartPostP\$vip} and \code{BartPostP\$within\_type\_vip}, respectively. We show the three variable importance measures for the object \code{BartPostP} in Figure~\ref{fig:bart}. As shown in the figure, the relevant importance rankings of the predictors are similar when different types of variable importance are used. However,  in the presence of mixed-type predictors, the permutation-based variable selection approach using either within-type VIP or MI improves the variable selection result of that using VIP, as shown in \cite{luo2021}.
%


\section{BNP Density Estimation with DPM Models} \label{sec:dpmm}

The DPM of multivariate normals model is a classic BNP model, which has been applied to density estimation in a variety of ways. The BART-DPM approach uses the approach of \cite{muller1996bayesian} to estimate the conditional distribution of the outcome given the estimated PS, by first modeling the joint density of the outcome and the estimated PS using the DPM of multivariate normals and then deriving the conditional density of interest from the joint density. In this section, we describe the DPM of multivariate normals model, the two MCMC samplers implemented in the \pkg{BNPqte} package and (joint and conditional) density estimation based on the two samplers. Simulated examples are also provided to illustrate the features of the \proglang{R} functions in the \pkg{BNPqte} package for density estimation.


\subsection{DPM of Multivariate Normals Model Specification} \label{sec:dpmm1}

We follow the notation of (\ref{eq:1}) and further assume that all the predictors are continuous. For every $i=1,\cdots,n$, let $\boldsymbol{z}_i = (Y_i, \boldsymbol{x}_i^{\top})^{\top}$. According to \cite{escobar1995bayesian} and \cite{muller1996bayesian}, the DPM of multivariate normals model on data $\{\boldsymbol{z}_i\}_{i=1}^n$ can be written as
\begin{align} \label{eq:6}
\boldsymbol{z}_i \,\vert\,\boldsymbol{\mu}_{i},\Sigma_{i} & \, \stackrel{\text{iid}}{\sim} \, \mathrm{Normal}\left(\boldsymbol{z}_i\,\vert\,\boldsymbol{\mu}_{i},\Sigma_{i}\right) \nonumber \\
\left(\boldsymbol{\mu}_{i},\Sigma_{i}\right)\,\vert\,G & \, \stackrel{\text{iid}}{\sim} \, G\\
G\,\vert\,\alpha,G_{0} & \, \sim \, \mathrm{DP}(\alpha G_{0}) \nonumber
\end{align}
where $\boldsymbol{\mu}_i$ is a $d$-dimensional ($d=p+1$) mean vector, $\Sigma_i$ is a $d \times d$ covariance matrix, $\alpha$ is the concentration parameter of the Dirichlet process (DP) prior $\mathrm{DP}(\alpha G_{0})$ and $G_0$ is the base distribution of the DP prior. The base distribution $G_0$ can be assumed as a conjugate normal-inverse-Wishart (NIW) distribution
\begin{equation} \label{eq:7}
G_{0} \, \equiv \, \mathrm{Normal}(\boldsymbol{\mu}\,\vert\,\boldsymbol{m},\frac{1}{\lambda}\Sigma)\times\mathrm{IW}(\Sigma\,\vert\,\nu,\Psi).
\end{equation}
Note that the inverse Wishart distribution $\mathrm{IW}(\Sigma\,\vert\,\nu,\Psi)$ is parameterized such that $\E(\Sigma) = \frac{\Psi}{\nu - d - 1}$. A hierarchical DPM of multivariate normals can be obtained by further assuming hyper-priors for the hyper-parameters $\boldsymbol{m}, \lambda, \Psi$ of $G_0$ and a prior for $\alpha$
\begin{align} \label{eq:8}
\alpha\,\vert\,a_{0},b_{0} & \, \sim \, \mathrm{Gamma}\left(\alpha\,\vert\,a_{0},b_{0}\right) \nonumber \\
\boldsymbol{m}\,\vert\,\boldsymbol{m}_{0},S_{0} & \, \sim \, \mathrm{Normal}\left(\boldsymbol{m}\,\vert\,\boldsymbol{m},S_{0}\right)\\
\lambda\,\vert\,\gamma_{1},\gamma_{2} & \, \sim \, \mathrm{Gamma}\left(\lambda\,\vert\,\gamma_{1},\gamma_{2}\right) \nonumber \\
\Psi\,\vert\,\nu_{0},\Psi_{0} & \, \sim \, \mathrm{Wishart}\left(\Psi\,\vert\,\nu_{0},\Psi_{0}\right) \nonumber
\end{align}
where $\mathrm{Gamma}\left(\alpha\,\vert\,a_{0},b_{0}\right)$ and $\mathrm{Wishart}\left(\Psi\,\vert\,\nu_{0},\Psi_{0}\right)$ are parameterized such that $\E(\alpha) = a_0/b_0$ and $\E(\Psi) = \nu_0 \Psi_0$, respectively. To complete the model specification, hyper-parameters ($\{\alpha, \boldsymbol{m}, \lambda, \nu, \Psi\}$ for Model~(\ref{eq:6}-\ref{eq:7}) and $\{a_0, b_0, \boldsymbol{m}_0, S_0, \gamma_1, \gamma_2, \nu, \nu_0, \Psi_0\}$ for Model~(\ref{eq:6}-\ref{eq:8})) need to be chosen. We choose these hyper-parameters according to \cite{taddy2008bayesian} and details can be found in Appendix~\ref{app:hyper}.


\subsection{Gibbs Sampler Based on Pólya Urn Scheme} \label{sec:neal}

The MCMC sampler of \cite{escobar1995bayesian} for the posterior distribution of the DPM model (\ref{eq:6}) iteratively draws each pair of $(\boldsymbol{\mu}_i, \Sigma_i)$ from its conditional distribution given both the data $\{\boldsymbol{z}_i\}_{i=1}^n$ and the other pairs of $\{(\boldsymbol{\mu}_j, \Sigma_j)\}_{j\neq i}$. This algorithm yields an ergodic Markov chain; however, it mixes poorly, because it updates each $(\boldsymbol{\mu}_i, \Sigma_i)$ one at a time and a value of $(\boldsymbol{\mu}_i, \Sigma_i)$ that is shared with many observations is hard to change.

Algorithm 8 of \cite{neal2000markov} solves this problem by first constructing an equivalent model of (\ref{eq:6}), which is the limit of a finite mixture model with $N$ components as $N$ goes to infinity. The finite mixture model has the following form
\begin{align} \label{eq:9}
\boldsymbol{z}_i\,\vert\,\kappa_{i},\phi & \, \stackrel{\text{iid}}{\sim} \, \mathrm{Normal}\left(\boldsymbol{z}_i\,\vert\,\boldsymbol{\zeta}_{\kappa_{i}},\Omega_{\kappa_{i}}\right) \nonumber \\
\kappa_{i}\,\vert\,\boldsymbol{\omega} & \, \stackrel{\text{iid}}{\sim} \, \sum\limits_{k=1}^{N}\omega_{i}\delta_{k}(\cdot) \nonumber \\
\boldsymbol{\omega}\,\vert\,\alpha & \, \sim \, \mathrm{Dirichlet}(\frac{\alpha}{N},\cdots,\frac{\alpha}{N})\\
\phi_{k}\stackrel{\triangle}{=}\left(\boldsymbol{\zeta}_{k},\Omega_{k}\right)\,\vert\,G & \, \stackrel{\text{iid}}{\sim} \, G \nonumber\\
G\,\vert\,\alpha,G_{0} & \, \sim \, \mathrm{DP}(\alpha G_{0}) \nonumber
\end{align}
where $\delta_k$ is a degenerate distribution at $k$, $\boldsymbol{\omega}$ is a set of $N$ mixing weights that sum to $1$, $\kappa_i$ is the latent cluster indicator for the observation $\boldsymbol{z}_i$, $\phi$ is the collection of all $\phi_k$'s and $(\boldsymbol{\zeta}_{\kappa_{i}},\Omega_{\kappa_{i}}) = (\boldsymbol{\mu}_i, \Sigma_i)$ for $1\leq k \leq N$ and $1\leq i \leq n$.

The Gibbs sampler of Algorithm 8 is applied to the limit of (\ref{eq:9}) with the mixing weights $\boldsymbol{\omega}$ integrated out, so the permanent state of the Markov chain only consists of $\{\kappa_i\}_{i=1}^n$ and $\phi_k$'s that are currently associated with some observations. Specifically, at each iteration of MCMC, they sample each $\kappa_i$ from the conditional distribution of $\kappa_i$ given $Y_i$, $\kappa_{-i}$ and $\phi$, where $\kappa_{-i} = \{\kappa_j\}_{j\neq i}$ and $\phi$ represents the set of $\phi_k$ currently associated with at least one observation. This conditional distribution can be obtained by taking the limit of the corresponding conditional distribution under the finite mixture model (\ref{eq:9}) and has the following form
\begin{align} \label{eq:10}
P(\kappa_{i}=k\,\vert\,\kappa_{-i},\boldsymbol{z}_{i},\phi) & \propto n_{-i,k}f(\boldsymbol{z}_{i}\,\vert\,\phi_k) & \qquad \text{if }k=\kappa_{j}\text{ for some }j\neq i, \nonumber \\
 & \propto\alpha\int f(\boldsymbol{z}_{i}\,\vert\,\phi)dG_{0}(\phi) & \qquad \text{if }k\neq\kappa_{j}\text{ for all }j\neq i,
\end{align}
where $n_{-i, k}$ is the number of $\kappa_j$ that are equal to $k$ for $j \neq i$ and $f(\boldsymbol{z}_{i}\,\vert\,\phi_k)$ is the normal likelihood. Since the conditional distribution of $\kappa_i$ in (\ref{eq:10}) has the form of the Pólya urn scheme, we refer to this sampler as the Pólya urn Gibbs sampler for convenience. To avoid potentially complex or intractable integration over $G_0$ in (\ref{eq:10}), Algorithm 8 introduces $m$ auxiliary parameters $\phi$ from $G_0$ with the idea that sampling $x$ from a distribution $\pi_x$ can be accomplished by sampling $(x,y)$ from some distributions $\pi_{xy}$ as long as the marginal distribution of $x$ from $\pi_{xy}$ is $\pi_x$. The auxiliary parameters are discarded after the update of $\kappa_i$. Once $\kappa_i$'s are updated, every $\phi_k \in \phi$ for the cluster $k$ is simply sampled from the posterior distribution of the prior $G_0$ and the likelihood of the observations $\{\boldsymbol{z}_{i}: \kappa_i=k, 1\leq i \leq n\}$ in that cluster. Algorithm 8 (with $m=1$) of \cite{neal2000markov} is restated in Algorithm \ref{alg:1} below and detailed computation for $\phi_k$'s and other hyper-parameters can be found in Appendix~\ref{app:neal}.

\begin{algorithm}
\caption{Algorithm 8 of \cite{neal2000markov} with $m=1$ auxiliary parameter}\label{alg:1}
\begin{algorithmic}[1] 

\Procedure{One Iteration of Markov Chain}{$\{\kappa_i\}_{i=1}^n, \{\phi_k\}_{k \in \{\kappa_i\}_{i=1}^n}$}
    \For{$i = 1, \cdots, n$}
        \State Let $k^{-}$ be the number of distinct $\kappa_j$ for $j \neq i$
        \State Label $\kappa_{-i}$ with values in $\{1,\cdots, k^{-}\}$
        \If{$\kappa_i = \kappa_j$ for some $j \neq i$}
            \State Draw $\phi_{k^{-}+1}$ from $G_0$
        \Else
            \State Set $\phi_{k^{-}+1} = \phi_{\kappa_i}$
        \EndIf
        \parState{Draw a new value for $\kappa_i$ from $\{1,\cdots, k^{-}+1\}$ with the following probabilities:
              $P(\kappa_{i}=k\,\vert\,\kappa_{-i},\boldsymbol{z}_{i},\phi_{1},\cdots,\phi_{k^{-}+1}) \propto
                \begin{cases}
                  n_{-i,k}f(\boldsymbol{z}_{i}\,\vert\,\phi_{k}) & \text{if }1\leq k\leq k^{-}\\
                  \alpha f(\boldsymbol{z}_{i}\,\vert\,\phi_{k})  & \text{if }k=k^{-}+1
                \end{cases}$
            }
    \EndFor
    \For{$k \in \{\kappa_1,\cdots,\kappa_n\}$}
        \State Draw a new value for $\phi_k$ from the posterior distribution $\phi_k \, \vert \, \{\boldsymbol{z}_i: \kappa_i = k\}$
    \EndFor
    \State Draw new values for $\alpha$ or hyper-parameters if they are assumed to be random
\EndProcedure

\end{algorithmic}
\end{algorithm}


\subsection{Blocked Gibbs Sampler Based on Truncated DPM} \label{sec:blocked}

The Pólya urn Gibbs sampler of \cite{neal2000markov} discussed above is the first sampler implemented in the \pkg{BNPqte} package and we refer to it as \code{neal} in the related \proglang{R} functions. We also implement the blocked Gibbs sampler of \cite{ishwaran2001gibbs}, which is referred to as \code{truncated} in the related \proglang{R} functions. The blocked Gibbs sampler is more efficient in density estimation than the Pólya urn Gibbs sampler, because it has the direct access to the full posterior distribution.

We first recall the stick-breaking representation of the Dirichlet process which can be expressed as
\begin{equation} \label{eq:11}
\mathcal{P}_{\infty}(\cdot) \,  =  \,  \sum\limits_{k=1}^{\infty} \omega_k \delta_{(\boldsymbol{\mu}_k, \Sigma_k)}(\cdot),
\end{equation}
where $\omega_1 = V_1$, $\omega_k = (1-V_1)\cdots (1-V_{k-1})V_k$ for $k \geq 2$ and $V_k \stackrel{\text{iid}}{\sim} \mathrm{Beta} (1, \alpha)$ for $k \geq 1$. Therefore, $G \, \vert \, \alpha, G_0 \, \sim \, \mathrm{DP}(\alpha G_0)$ in  model~(\ref{eq:6}) can be rewritten as $G \, \sim \, \mathcal{P}_{\infty}$. 

Based on the representation (\ref{eq:11}), \cite{ishwaran2000markov} and \cite{ishwaran2001gibbs} find a random probability measure, denoted by $\mathcal{P}_N$, almost surely converging to $\mathcal{P}_{\infty}$. The random probability measure $\mathcal{P}_N$ is obtained by truncating the higher-order terms in the representation (\ref{eq:11}) and has the following form
\begin{equation} \label{eq:12}
\mathcal{P}_N(\cdot) \,  =  \,  \sum\limits_{k=1}^{N} \omega_k \delta_{(\boldsymbol{\mu}_k, \Sigma_k)}(\cdot),
\end{equation}
where
\begin{align} \label{eq:13}
\omega_1 & \, = \, V_1, \qquad \omega_k \, = \, V_k \prod\limits_{j=1}^{k-1} (1-V_j) \quad \text{for } 2 \leq k \leq N, \\
V_N & \, =  \, 1, \qquad \, \, V_k  \, \stackrel{\text{iid}}{\sim} \, \mathrm{Beta}(1, \alpha) \quad \text{for } 1 \leq k \leq N-1. \nonumber
\end{align}
The weights $\boldsymbol{\omega} = \{\omega_k\}_{k=1}^N$ constructed by (\ref{eq:13}) essentially follow a generalized Dirichlet distribution $\mathrm{GD}(1,\alpha,\cdots,1,\alpha)$.

They then build an efficient MCMC algorithm for the DPM model~(\ref{eq:6}) with the DP prior $\mathcal{P}_{\infty}$ replaced by $\mathcal{P}_N$, i.e., the truncated DPM model. Similar to Algorithm 8 of \cite{neal2000markov}, they also include latent cluster indicators and the truncated DPM model can be expressed as follows
\begin{align} \label{eq:14}
\boldsymbol{z}_i\,\vert\,\kappa_{i},\phi & \, \stackrel{\text{iid}}{\sim} \, \mathrm{Normal}\left(\boldsymbol{z}_i\,\vert\,\boldsymbol{\zeta}_{\kappa_{i}},\Omega_{\kappa_{i}}\right) \nonumber \\
\kappa_{i}\,\vert\,\boldsymbol{\omega} & \, \stackrel{\text{iid}}{\sim} \, \sum\limits_{k=1}^{N}\omega_{i}\delta_{k}(\cdot) \\
\boldsymbol{\omega}\,\vert\,\alpha & \, \sim \, \mathrm{GD}(1,\alpha,\cdots,1,\alpha) \nonumber \\
\phi_{k}\stackrel{\triangle}{=}\left(\boldsymbol{\zeta}_{k},\Omega_{k}\right) & \, \stackrel{\text{iid}}{\sim} \, G_0. \nonumber
\end{align}
Model~(\ref{eq:14}) allows block updates for all the parameters in a Gibbs sampler, which leads to rapid mixing of the Markov chain. Also, there is no need to integrate any parameters out, which permits direct inference for the posterior $\mathcal{P}_N \, \vert \, \{\boldsymbol{z}_i\}_{i=1}^n$.

The blocked Gibbs sampler for the truncated DPM model~(\ref{eq:14}) is based on the following form of the full posterior distribution
\begin{align} \label{eq:15}
f(\{\phi_k\}_{k=1}^N, \{\kappa_i\}_{i=1}^n, \boldsymbol{\omega} \, \vert \, \{\boldsymbol{z}_i\}_{i=1}^n, \alpha) \, \propto \,  & \left[ \prod\limits_{j=1}^{K^*} f(\phi_{\kappa_j^*}) \prod\limits_{\{i:\kappa_i = \kappa_j^*\}} f(\boldsymbol{z}_i \, \vert \, \phi_{\kappa_j^*}) \right] \times \left[ \prod\limits_{k \notin \boldsymbol{\kappa}^*} f(\phi_k) \right] \nonumber \\
&  \times \left[ \prod\limits_{i=1}^n f(\kappa_i \, \vert \, \boldsymbol{\omega}) \right] \times f(\boldsymbol{\omega} \, \vert \, \alpha),
\end{align}
where $\boldsymbol{\kappa}^* = \{\kappa_j^*\}_{j=1}^{K^*}$ is the collection of unique values in $\{\kappa_i\}_{i=1}^n$. Hyper-priors can also be included in (\ref{eq:15}) if the hierarchical structure (\ref{eq:8}) is assumed. The algorithm for the truncated DPM model~(\ref{eq:14}) is summarized in Algorithm \ref{alg:2} and details of the posterior computation can be found in Appendix~\ref{app:truncated}.

The posterior samples obtained from Algorithm \ref{alg:2} are used for approximating posterior samples of the DPM model (\ref{eq:6}). To achieve an adequate approximation of the Dirichlet process, we need to choose a reasonably large $N$. Theorem 2 of \cite{ishwaran2000markov} provides guidance to select $N$. In the \pkg{BNPqte} package, we set $N=50$ by default.

\begin{algorithm}
\caption{Blocked Gibbs Sample of \cite{ishwaran2001gibbs}}\label{alg:2}
\begin{algorithmic}[1]

\Procedure{One Iteration of Markov Chain}{$\{\kappa_i\}_{i=1}^n, \{\phi_k\}_{k=1}^N, \{\omega_k\}_{k=1}^N$}
    \For{$k = 1, \cdots, N$}
        \If{$k \in \boldsymbol{\kappa}^*$}
            \State Sample a new value for $\phi_k$ from the posterior distribution $\phi_k \, \vert \, \{\boldsymbol{z}_i: \kappa_i = k \}$
        \Else
            \State Sample a new value for $\phi_k$ from the prior distribution of $\phi_k$
        \EndIf
    \EndFor
    \State Sample a vector of new values for $\{\omega_k\}_{k=1}^N$ from $\mathrm{GD}(a_1,b_1,\cdots,a_{N-1},b_{N-1})$ with 
    \begin{equation*}
    a_k = n_k + 1 \quad \text{ and } \quad b_k = N_k + \alpha,
    \end{equation*}
    ~~~~where  $n_k = \vert \{i: \kappa_i = k \}\vert$ for $1 \leq k \leq N$ and $N_k = n_{k+1}+\cdots,n_N$ for $1 \leq k \leq N-1$
    \For{$i = 1, \cdots, n$}
        \State Sample a new value for $\kappa_i$ from $\{1,\cdots, N\}$ with probabilities proportional to $$(\omega_1 f(\boldsymbol{z}_i \, \vert \, \phi_1), \cdots, \omega_N f(\boldsymbol{z}_i \, \vert \, \phi_N))$$
    \EndFor
    \State Draw new values for $\alpha$ or hyper-parameters if they are assumed to be random
\EndProcedure

\end{algorithmic}
\end{algorithm}


\subsection{Density Estimation Based on Blocked Gibbs Sampler} \label{sec:density}

The DPM of multivariate normals model can be used to estimate the density for continuous data. Consider the problem of estimating the predictive density for a future observation $\boldsymbol{z}_{n+1}$ given the observations $\{\boldsymbol{z}_i\}_{i=1}^n$. Let $f(\boldsymbol{z}_{n+1} \, \vert \, \{\boldsymbol{z}_i\}_{i=1}^n)$ be the predictive density of $\boldsymbol{z}_{n+1}$ given the observations $\{\boldsymbol{z}_i\}_{i=1}^n$ under the DPM of multivariate normals model. Based on the stick-breaking representation of the DP prior in (\ref{eq:11}), we have that
\begin{equation} \label{eq:16}
f(\boldsymbol{z}_{n+1} \, \vert \, \{\boldsymbol{z}_i\}_{i=1}^n) \, = \, \int \int f(\boldsymbol{z}_{n+1} \, \vert \, \boldsymbol{\mu}_{n+1}, \Sigma_{n+1}) dF(\boldsymbol{\mu}_{n+1}, \Sigma_{n+1} \, \vert \, \mathcal{P}_{\infty}) dF(\mathcal{P}_{\infty} \, \vert \, \{\boldsymbol{z}_i\}_{i=1}^n),
\end{equation}
where $f(\boldsymbol{z}_{n+1} \, \vert \, \boldsymbol{\mu}_{n+1}, \Sigma_{n+1})$ is the assumed normal likelihood and $F(\mathcal{P}_{\infty} \, \vert \, \{\boldsymbol{z}_i\}_{i=1}^n)$ in the outer integral is the posterior distribution of the Dirichlet process.

When the truncated DPM model (\ref{eq:14}) is used to approximate the exact DPM model (\ref{eq:6}), the inside integral in (\ref{eq:16}) can be estimated by
\begin{equation} \label{eq:17}
\sum\limits_{k=1}^N \omega_k^{\{l\}} f(\boldsymbol{z}_{n+1} \, \vert \, \boldsymbol{\zeta}_k^{\{l\}}, \Omega_k^{\{l\}}),
\end{equation}
where $\{\omega_k^{\{l\}}, \boldsymbol{\zeta}_k^{\{l\}}, \Omega_k^{\{l\}} \}_{k=1}^N$ is the $l^{\text{th}}$  posterior sample from MCMC. The predictive density $f(\boldsymbol{z}_{n+1} \, \vert \, \{\boldsymbol{z}_i\}_{i=1}^n)$ can then be estimated by the average of (\ref{eq:17}) across all the MCMC samples.

We can also estimate the predictive conditional density of $Y_{n+1}$ given $\boldsymbol{x}_{n+1}$ based on the DPM model (\ref{eq:6}) on the joint data $\{\boldsymbol{z}_i = (Y_{i}, \boldsymbol{x}_{i}^{\top})^{\top} \}_{i=1}^n$. Let $f(Y_{n+1} \, \vert \, \boldsymbol{x}_{n+1},  \{\boldsymbol{z}_i\}_{i=1}^n)$ be the predictive conditional density of interest under the DPM model. It can be directly derived from (\ref{eq:16}) and has the following form
\begin{equation} \label{eq:18}
f(Y_{n+1} \, \vert \, \boldsymbol{x}_{n+1},  \{\boldsymbol{z}_i\}_{i=1}^n)  \, = \, \int \frac{f(Y_{n+1}, \boldsymbol{x}_{n+1} \, \vert \, \mathcal{P}_{\infty})}{f(\boldsymbol{x}_{n+1} \, \vert \, \mathcal{P}_{\infty})} dF(\mathcal{P}_{\infty} \, \vert \,  \{\boldsymbol{z}_i\}_{i=1}^n).
\end{equation}
The numerator of the fraction in (\ref{eq:18}) is the inside integral in (\ref{eq:16}), so the estimation is simply (\ref{eq:17}) if the blocked Gibbs sampler is used. The denominator of the fraction is the marginal density of $\boldsymbol{x}_{n+1}$, which can be obtained by integrating the numerator over $Y_{n+1}$, so it can be estimated by the integral of (\ref{eq:17}) over $Y_{n+1}$. In conclusion, the integrand in (\ref{eq:18}) can be estimated by
\begin{equation} \label{eq:19}
\sum\limits_{k=1}^N \omega_{k}^{\{l\}}(\boldsymbol{x}_{n+1}) f_{\mathrm{Normal}}(Y_{n+1} \, \vert \, \beta_{0k}^{\{l\}} + \boldsymbol{x}_{n+1}^{\top}\boldsymbol{\beta}_{k}^{\{l\}}, \sigma_{k}^{2\{l\}}),
\end{equation}
where $f_{\mathrm{Normal}}(\cdot \, \vert \, \beta, \sigma^2)$ represents a univariate normal density with mean $\beta$ and variance $\sigma^2$,
\begin{align} \label{eq:20}
& \boldsymbol{\beta}_{k}^{\{l\}} \, = \, \Omega_{12k}^{\{l\}} \Omega_{22k}^{\{l\}-1}, \quad \beta_{0k}^{\{l\}} \, = \, \zeta_{1k}^{\{l\}} - \boldsymbol{\beta}_{k}^{\{l\}} \boldsymbol{\zeta}_{2k}^{\{l\}}, \quad \sigma_{k}^{2\{l\}}  \, = \, \Omega_{11k}^{\{l\}} - \boldsymbol{\beta}_{k}^{\{l\}} \Omega_{21k}^{\{l\}}, \nonumber \\
& \omega_{k}^{\{l\}}(\boldsymbol{x}_{n+1}) \, = \, \frac{\omega_k^{\{l\}} f(\boldsymbol{x}_{n+1} \, \vert \, \boldsymbol{\zeta}_{2k}^{\{l\}}, \Omega_{22k}^{\{l\}})}{\sum\limits_{j=1}^N \omega_j^{\{l\}} f(\boldsymbol{x}_{n+1} \, \vert \, \boldsymbol{\zeta}_{2j}^{\{l\}}, \Omega_{22j}^{\{l\}})},  \\
& \boldsymbol{\zeta}_{k}^{\{l\}} \,=\,\left(\begin{array}{c}
\zeta_{1k}^{\{l\}}\\
\boldsymbol{\zeta}_{2k}^{\{l\}}
\end{array}\right) \quad \text{and} \quad \Omega_{k}^{\{l\}}\,=\,\left(\begin{array}{cc}
\Omega_{11k}^{\{l\}} & \Omega_{12k}^{\{l\}}\\
\Omega_{21k}^{\{l\}} & \Omega_{22k}^{\{l\}}
\end{array}\right). \nonumber
\end{align}
The estimator in (\ref{eq:19}) induces an estimator for the predictive conditional mean of $Y$ given a future $\boldsymbol{x}_{n+1}$, which can be expressed as $\hat{\E}(Y \, \vert \, \boldsymbol{x}_{n+1}) = \sum\limits_{k=1}^N \omega_{k}^{\{l\}}(\boldsymbol{x}_{n+1}) (\beta_{0k}^{\{l\}} + \boldsymbol{x}_{n+1}^{\top}\boldsymbol{\beta}_{k}^{\{l\}})$. The predictive conditional CDF of $Y_{n+1}$ given $\boldsymbol{x}_{n+1}$ can be similarly estimated by an estimator which has the form of (\ref{eq:19}) with the normal density $f_{\mathrm{Normal}}$ replaced by the corresponding normal CDF.

Density estimation based on the Pólya urn Gibbs sampler is more complicated than that based on the blocked Gibbs sampler, because it samples from the marginal posterior distribution and does not have the access to the mixing weights $\omega_k$'s. We implement the approach of \cite{muller1996bayesian}, when the Pólya urn Gibbs sampler is employed for the MCMC sampling. We describe the details in Appendix~\ref{app:estimation}.


\subsection{Illustrations}

The \pkg{BNPqte} package implements joint density estimation by the \proglang{R} function \fct{DPMdensity} and conditional density estimation by the \proglang{R} function \fct{DPMcdensity}. Arguments of the two functions can be divided into five components: data inputs, prediction inputs, model specification, MCMC parameters and diagnostics setting, the last three of which are shared between the two functions. We first introduce the three shared components. 

In the model specification part, two important arguments are \code{updateAlpha} and \code{useHyperpriors}, which are described below.
\begin{itemize}
    \item \code{updateAlpha = TRUE} implies that the concentration parameter $\alpha$ is assumed to be random as (\ref{eq:8}), otherwise it is assumed to be a fixed value
    \item \code{useHyperpriors = TRUE} implies that the hyper-parameters of the base distribution $G_0$ are assumed to be random as (\ref{eq:8}), otherwise they are assumed to be fixed values
\end{itemize}
By setting these two arguments to different values, both hierarchical and non-hierarchical DPM models can be fit in our package. Other arguments for specifying hyper-parameters can often be omitted since the default values are adequate for most purposes. 

Arguments for MCMC parameters include those for specifying the sampling algorithm and those for setting the number of posterior samples burned, kept and thinned. We list these arguments below.
\begin{itemize}
    \item \code{method} is a character string; the Pólya urn Gibbs sampler is chosen if \code{method = "neal"} and the blocked Gibbs sampler is chosen if \code{method = "truncated"}
    \item \code{nclusters} is the number of clusters $N$ pre-specified for the blocked Gibbs sampler if \code{method = "truncated"}
    \item \code{nskip}, \code{ndpost} and \code{keepevery} are the number of posterior samples burned, kept and thinned, respectively
\end{itemize}

The last part of arguments shared between \fct{DPMdensity} and \fct{DPMcdensity} includes an argument \code{diag} of boolean type. By setting \code{diag = TRUE}, logarithm likelihood and logarithm marginal partition posterior (only for the blocked Gibbs sampler) are calculated at each iteration of the Markov chain and are returned to users for further diagnostics analysis. We discuss this in Section~\ref{sec:diagnostics}. 

Unlike the three parts of arguments discussed above, arguments for data and prediction are different for \fct{DPMdensity} and \fct{DPMcdensity} and they are either always or frequently needed from users. The data and prediction inputs of \fct{DPMdensity} are as follows.
\begin{itemize}
    \item \code{y} is a matrix giving the data from which the density estimate is to be computed
    \item \code{ngrid} and \code{grid} produce the grid points where the density estimate is evaluated; they are only used when \code{y} is bivariate and nothing is evaluated when \code{ngrid = 0} and \code{grid = NULL}
\end{itemize}

The data and prediction inputs of \fct{DPMcdensity} are slightly different from above, as shown below.
\begin{itemize}
    \item \code{y} is a vector giving the values for which the conditional density (and/or conditional mean and/or conditional CDF) estimate is to be computed
    \item \code{x} is a vector or matrix specifying the predictors
    \item \code{ngrid} and \code{grid} produce the grid points where the conditional density (and/or conditional mean and/or conditional CDF) estimate is evaluated; nothing is evaluated when \code{ngrid = 0} and \code{grid = NULL}
    \item \code{xpred} is a vector or matrix giving the values of the predictors for prediction
    \item \code{type.pred} is a vector of strings; the conditional density (and/or conditional mean and/or conditional CDF) is estimated if \code{"pdf"} (and/or \code{"meanReg"} and/or \code{"cdf"}) is included in \code{type.pred}
    \item \code{compute.band} is a boolean argument indicating whether a credible band for the conditional density (and/or conditional mean and/or conditional CDF) is computed and \code{type.band} is a string indicating the type of credible bands to be computed; $95\%$ pointwise highest posterior interval (or Bayesian credible interval) (\cite{chen1999monte}) is computed if \code{type.band = "HPD"} (or \code{type.band = "BCI"})
\end{itemize}

The result of running \fct{DPMdensity} (or \fct{DPMcdensity}) is returned in an object of class \code{"DPMdensity"} (or \code{"DPMcdensity"}) which essentially is a list. Besides the prediction results, the posterior samples are stored in a sub-list called \code{posterior} and the latest posterior sample is stored in a sub-list called \code{state} which can be used as initial values for a new Markov chain. We illustrate these outputs with simulated examples later.

We also provide S3 method \fct{predict} and \fct{plot} functions for classes \code{"DPMdensity"} and \code{"DPMcdensity"}. Users can use \fct{predict} to obtain the prediction of the grid points provided by them and use \fct{plot} to visualize the prediction results from objects of class \code{"DPMdensity"} and \code{"DPMcdensity"}.

Next, we illustrate the usage of the \proglang{R} functions \fct{DPMdensity} and \fct{DPMcdensity} with two simulated examples.

\subsubsection{Joint density estimation}

We simulate $n=500$ independent and identically distributed (i.i.d.) observations $\{\boldsymbol{y}_i\}_{i=1}^n$ from a mixture of three bivariate normal distributions with equal weights. The mean vectors of the three normals are $\boldsymbol{\zeta}_{01}^{\top} = (2, -1)$, $\boldsymbol{\zeta}_{02}^{\top} = (1, 0)$ and $\boldsymbol{\zeta}_{03}^{\top} = (-1, -1)$, respectively; the three normals share the same covariance matrix which is a $2$ by $2$ diagonal matrix with diagonal elements equal to $0.5$, i.e., $\Sigma_0 = 0.5\mathbb{I}_{2}$. The true density function of $\boldsymbol{y}$ is given below
\begin{equation} \label{exp:2}
f(\boldsymbol{y}) \, = \, \frac{1}{3} f_{\mathrm{Normal}} (\boldsymbol{y} \, \vert \, \boldsymbol{\zeta}_{01}, \Sigma_0) + \frac{1}{3} f_{\mathrm{Normal}} (\boldsymbol{y} \, \vert \, \boldsymbol{\zeta}_{02}, \Sigma_0) + \frac{1}{3} f_{\mathrm{Normal}} (\boldsymbol{y} \, \vert \, \boldsymbol{\zeta}_{03}, \Sigma_0).
\end{equation}
\pkg{BNPqte} provides the \proglang{R} function \fct{ThreeNormals} to generate the data described above. As shown below, besides the observations (i.e., \code{$y}), the true density function (i.e., \code{$dtrue}) is also returned by the function.
\begin{Schunk}
\begin{Sinput}
R> set.seed(0)
R> DpmData1 = ThreeNormals(n = 500)
R> names(DpmData1)
\end{Sinput}
\begin{Soutput}
[1] "y"     "dtrue"
\end{Soutput}
\end{Schunk}
We model the joint density of $\boldsymbol{y}$ by fitting a hierarchical DPM of multivariate normals model on the observations $\{\boldsymbol{y}_i\}_{i=1}^n$. Specifically, we run the \proglang{R} function \fct{DPMdensity} on the data \code{DpmData1\$y}. The Pólya urn Gibbs sampler (i.e., \code{method = "neal"}) and the blocked Gibbs sampler (i.e., \code{method = "truncated"}) are used, respectively, and the corresponding results are returned in the objects \code{JtDpmP} and \code{JtDpmB}. We burn in the first $5000$ posterior samples (i.e., \code{nskip = 5000}) and then save $5000$ posterior samples (i.e., \code{ndpost = 5000}) by keeping every 3rd sample (i.e., \code{keepevery = 3}). The estimated density is evaluated at $1000$ grid points (i.e., \code{ngrid = 1000}) generated from the observed data.
\begin{Schunk}
\begin{Sinput}
R> t1 = Sys.time()       
R> JtDpmB = DPMdensity(y = DpmData1$y, ngrid = 1000, 
+                      method = "truncated", nclusters = 50, 
+                      nskip = 5000, ndpost = 5000, keepevery = 3)
R> t2 = Sys.time()
R> JtDpmP = DPMdensity(y = DpmData1$y, ngrid = 1000, method = "neal",
+                      nskip = 5000, ndpost = 5000, keepevery = 3) 
R> t3 = Sys.time()
\end{Sinput}
\end{Schunk}
The returned object \code{JtDpmB} is a list and has the following components.
\begin{Schunk}
\begin{Sinput}
R> names(JtDpmB) 
\end{Sinput}
\begin{Soutput}
 [1] "method"          "updateAlpha"     "useHyperpriors" 
 [4] "status"          "state"           "posterior"      
 [7] "prediction"      "predict.pdfs"    "predict.pdf.avg"
[10] "grid1"           "grid2"           "proc.time"      
\end{Soutput}
\end{Schunk}
Posterior samples of (hyper-)parameters are returned in the component \code{$posterior}, as follows.
\begin{Schunk}
\begin{Sinput}
R> names(JtDpmB$posterior) 
\end{Sinput}
\begin{Soutput}
[1] "Zeta"   "Omega"  "lw"     "kappa"  "alpha"  "m"      "lambda"
[8] "Psi"   
\end{Soutput}
\end{Schunk}
Prediction results are returned in the components \code{$predict.pdfs} and \code{$predict.pdf.avg}. The component \code{$predict.pdfs} is a list of length \code{ndpost}. Each element of the list is a matrix of the estimated density evaluated at grid points (\code{$grid1}, \code{$grid2}), using a single MCMC sample. The component \code{$predict.pdf.avg} is the averaged \code{$predict.pdfs} across the MCMC samples.

\begin{figure}[h!]
\centering
\includegraphics{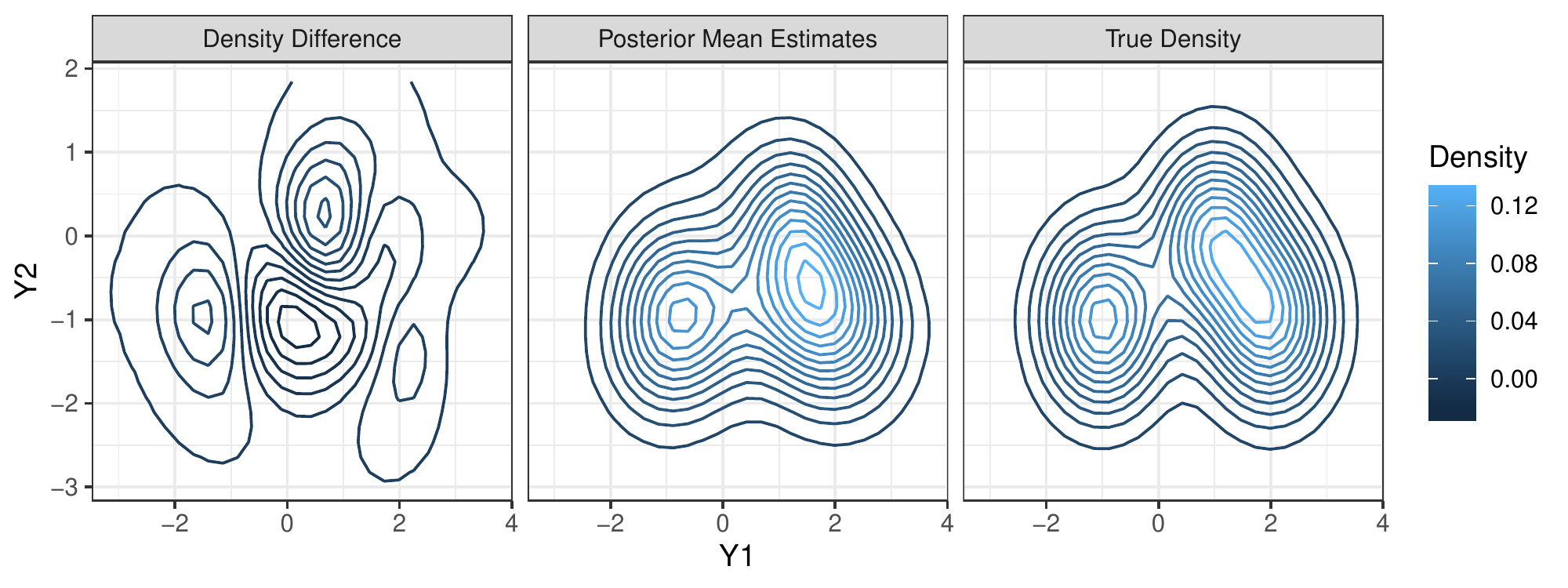}
\caption{\label{fig:jt1}From left to right, subfigures are contours of the difference between the estimated density and the true density, the estimated density, and the true density, respectively, for the DPM of multivariate normals model using the blocked Gibbs sampler for the example in (\ref{exp:2}).}
\end{figure}

We can visualize the prediction results by applying the \fct{plot} function to the object \code{JtDpmB}, as shown in Figure~\ref{fig:jt1}. Note that if the \code{true_density_fun} is not provided to the \fct{plot} function, only the posterior mean estimate is plotted.
\begin{Schunk}
\begin{Sinput}
R> plot(JtDpmB, diff = TRUE, true_density_fun = DpmData1$dtrue, 
+       type.plot="contour")
\end{Sinput}
\end{Schunk}
The DPM model using the Pólya urn Gibbs sampler has similar prediction results, as shown in Figure~\ref{fig:jt2} in Appendix~\ref{app:simulation}. The \fct{plot} function in \pkg{BNPqte} also provides \pkg{plotly} 3D-surface (HTML output) for the estimated density from \fct{DPMdensity}. Users can obtain the 3D-surface by setting \code{type.plot = "surface"}, if working in a HTML environment.

The run times of the blocked Gibbs sampler and the Pólya urn Gibbs sampler are similar, when the DPM model is applied to the joint density estimation.
\begin{Schunk}
\begin{Sinput}
R> tB = difftime(t2, t1)
R> tP = difftime(t3, t2)
R> cat("Run time of Blocked Gibbs Sampler: ", round(tB, 2), units(tB), "\n",
+      "Run time of Pólya urn Sampler: ", round(tP, 2), units(tP), sep = "")
\end{Sinput}
\begin{Soutput}
Run time of Blocked Gibbs Sampler: 1.19mins
Run time of Pólya urn Sampler: 1.1mins
\end{Soutput}
\end{Schunk}

With a fitted object like \code{JtDpmB} and \code{JtDpmP}, users can evaluate the estimated density at some given grid points by plugging the object in the \fct{predict} function. The returned object from the \fct{predict} function is also of class \code{"DPMdensity"}, so the prediction result can be viewed by the \fct{plot} function as well. We show the syntax of the \fct{predict} function below.
\begin{Schunk}
\begin{Sinput}
R> grids = matrix(runif(10), ncol = 2)
R> predict(object = JtDpmB, grid = grids)
\end{Sinput}
\end{Schunk}

\subsubsection{Conditional density estimation}

\begin{figure}[h!]
\centering
\includegraphics{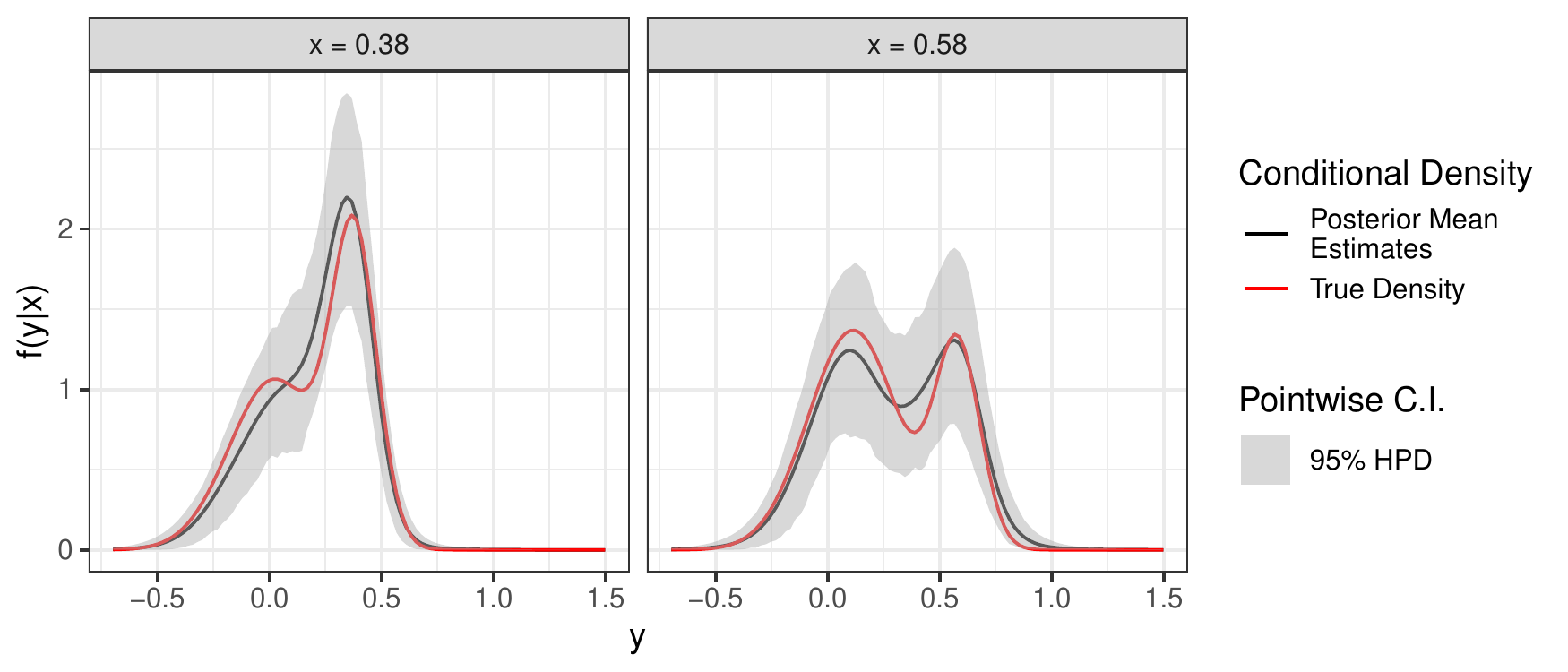}
\caption{\label{fig:cd1}Estimated conditional density from the DPM of multivariate normals model using the blocked Gibbs sampler for the example in (\ref{exp:3}).}
\end{figure}

We show the usage of the \proglang{R} function \fct{DPMcdensity} with a simulated example from \cite{dunson2007bayesian}, which is also replicated by the \pkg{DPpackage} \proglang{R} package. We first sample $n=500$ i.i.d. covariate values $\{x_i\}_{i=1}^n$ from $\mathrm{Uniform}(0,1)$, and then for each $x_i$, generate the response $y_i$ from a mixture of a normal linear regression model and a normal non-linear regression model, with the mixture weights depending on the predictor $x_i$. The true conditional density of $y$ given $x$ can be expressed as follows:
\begin{equation} \label{exp:3}
f(y \, \vert \, x) \, = \, \exp(-2x) f_{\mathrm{Normal}}(y \, \vert \, x, 0.01) + (1 - \exp(-2x)) f_{\mathrm{Normal}}(y \, \vert \, x^4, 0.04).
\end{equation}
The data described above can be generated by the \proglang{R} function \fct{DunsonExample} in our package. Besides the data (i.e., \code{$y} and \code{$x}), the density function (i.e., \code{$dtrue}), mean (i.e., \code{$mtrue}) and CDF (i.e., \code{$ptrue}) of the true conditional distribution are also returned by the function, as shown below.
\begin{Schunk}
\begin{Sinput}
R> set.seed(0)
R> DpmData2 = DunsonExample(n = 500)
R> names(DpmData2)
\end{Sinput}
\begin{Soutput}
[1] "y"     "x"     "mtrue" "dtrue" "ptrue"
\end{Soutput}
\end{Schunk}
We model the conditional density of $y$ given $x$ by fitting a hierarchical DPM of multivariate normals model on the joint data $\{y_i, x_i\}_{i=1}^n$  and then calculating the conditional density from the estimated joint density, as described in Section~\ref{sec:density} and Appendix~\ref{app:estimation}. Specifically, we run the function \fct{DPMcdensity} twice with the Pólya urn Gibbs sampler and the blocked Gibbs sampler, respectively. The corresponding returned objects are \code{CdDpmP} and \code{CdDpmB}, which are of class \code{"DPMcdensity"}. MCMC settings are the same as those in the joint density estimation above. Some arguments set to their default values are omitted in the following code chunk.
\begin{Schunk}
\begin{Sinput}
R> type.pred = c("pdf", "meanReg", "cdf")
R> xpred = seq(0, 1, 0.02)
R> t4 = Sys.time()       
R> CdDpmB = DPMcdensity(y = DpmData2$y, x = DpmData2$x, 
+                       xpred = xpred, ngrid = 100, type.pred = type.pred,
+                       nskip = 5000, ndpost = 5000, keepevery = 3)
R> t5 = Sys.time()
R> CdDpmP = DPMcdensity(y = DpmData2$y, x = DpmData2$x, method = "neal",
+                       xpred = xpred, ngrid = 100, type.pred = type.pred,
+                       nskip = 5000, ndpost = 5000, keepevery = 3)
R> t6 = Sys.time()
\end{Sinput}
\end{Schunk}
The returned object \code{CdDpmB} has similar components to the object \code{JtDpmB} obtained in joint density estimation, except those about prediction results. The component \code{$predict.pdfs} consists of the estimated conditional density at each MCMC iteration, and \code{$predict.pdf.avg} is the average of \code{$predict.pdfs} across MCMC iterations. The components \code{$predict.pdf.lower} and \code{$predict.pdf.upper} are the lower and upper bounds of the pointwise $95\%$ credible intervals of \code{$predict.pdfs}. The estimated conditional mean and CDF have similar names of return values as the estimated conditional density, except that the \code{pdf} in the name of a return value for the estimated conditional density is replaced by \code{meanReg} (or \code{CDF}) for the estimated conditional mean (or CDF), as shown below.
\begin{Schunk}
\begin{Sinput}
R> names(CdDpmB)
\end{Sinput}
\begin{Soutput}
 [1] "updateAlpha"           "useHyperpriors"       
 [3] "status"                "state"                
 [5] "posterior"             "predict.meanRegs"     
 [7] "predict.meanReg.avg"   "predict.meanReg.lower"
 [9] "predict.meanReg.upper" "predict.pdfs"         
[11] "predict.pdf.avg"       "predict.pdf.lower"    
[13] "predict.pdf.upper"     "predict.cdfs"         
[15] "predict.cdf.avg"       "predict.cdf.lower"    
[17] "predict.cdf.upper"     "proc.time"            
[19] "prediction"            "type.pred"            
[21] "compute.band"          "type.band"            
[23] "xpred"                 "grid"                 
\end{Soutput}
\end{Schunk}
Like the function \fct{DPMdensity}, we can also visualize the prediction results of the function \fct{DPMcdensity} via the function \fct{plot} as below. The prediction results of the DPM model using the blocked Gibbs sampler are shown in Figure~\ref{fig:cd1}, Figure~\ref{fig:cd2} and Figure~\ref{fig:cd3} and those using the Pólya urn Gibbs sampler can be found in Appendix~\ref{app:simulation}. The two Gibbs samplers produce similar prediction results.
\begin{Schunk}
\begin{Sinput}
R> plot(CdDpmB, xpred.idx = c(20, 30), true_pdf_fun = DpmData2$dtrue,
+       true_cdf_fun = DpmData2$ptrue, true_meanReg_fun = DpmData2$mtrue)
\end{Sinput}
\end{Schunk}
\begin{figure}[h!]
\centering
\includegraphics{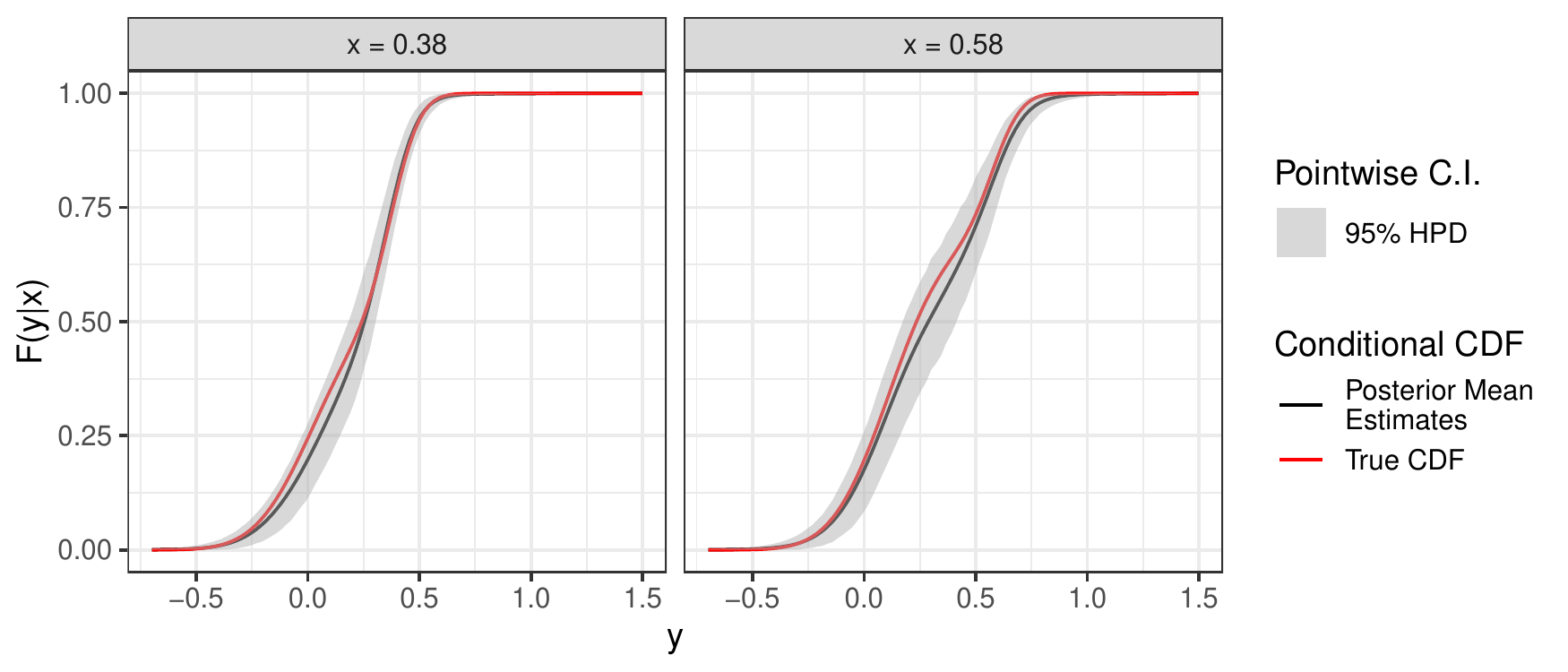}
\caption{\label{fig:cd2}Estimated conditional CDF from the DPM of multivariate normals model using the blocked Gibbs sampler for the example in (\ref{exp:3}).}
\end{figure}
\begin{figure}[h!]
\centering
\includegraphics{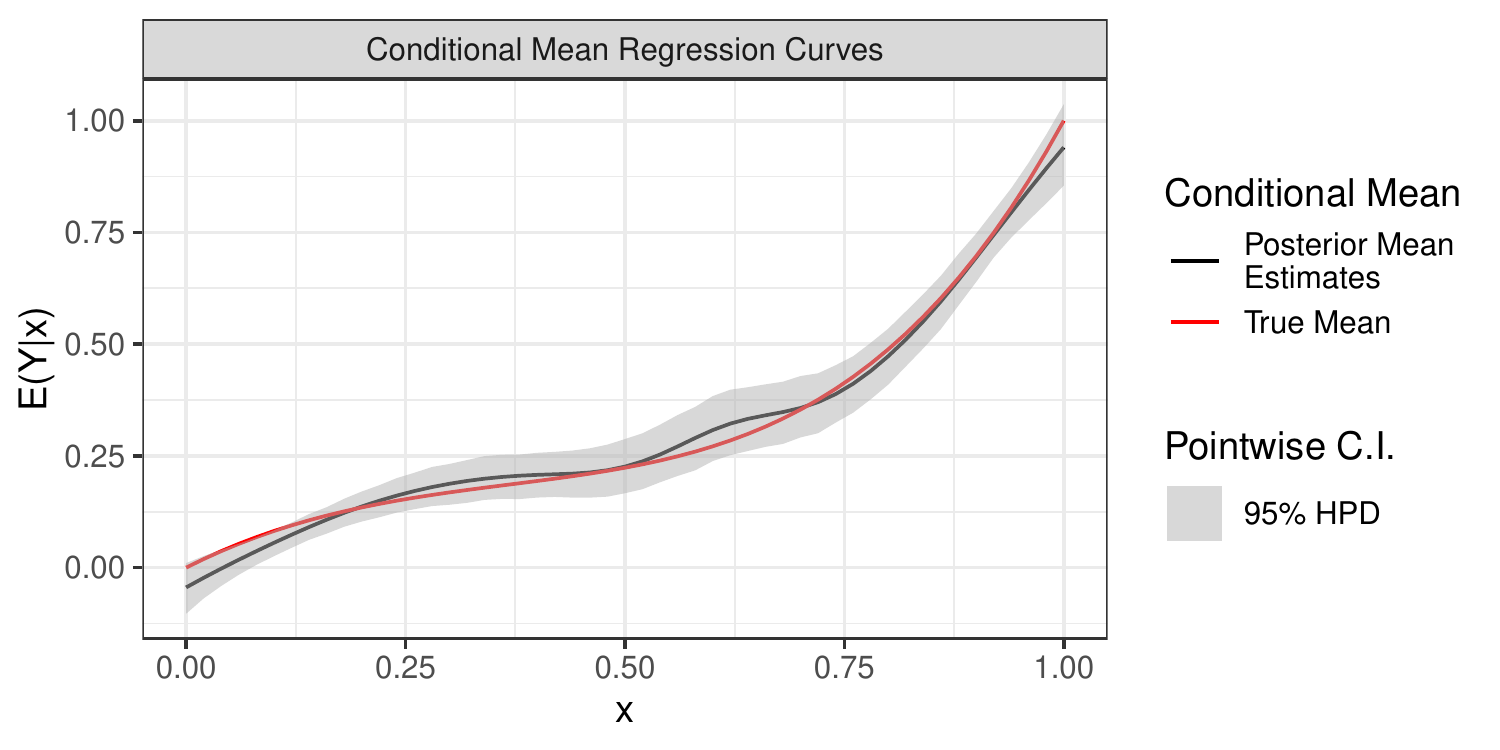}
\caption{\label{fig:cd3}Estimated conditional mean from the DPM of multivariate normals model using the blocked Gibbs sampler for the example in (\ref{exp:3}).}
\end{figure}
However, the run time of the function \fct{DPMcdensity} using the Pólya urn Gibbs sampler is much longer ($7$ times longer in this example) than that using the blocked Gibbs sampler. This is because the Pólya urn Gibbs sampler does not have the direct access to the full posterior distribution and requires a complicated algorithm to obtain it. We discuss this in detail in Appendix~\ref{app:estimation}.
\begin{Schunk}
\begin{Sinput}
R> ctB = difftime(t5, t4)
R> ctP = difftime(t6, t5)
R> cat("Run time of blocked Gibbs Sampler: ", round(ctB, 2), units(ctB), "\n",
+      "Run time of Pólya urn Sampler: ", round(ctP, 2), units(ctP), sep = "")
\end{Sinput}
\begin{Soutput}
Run time of blocked Gibbs Sampler: 3.19mins
Run time of Pólya urn Sampler: 23.96mins
\end{Soutput}
\end{Schunk}
When using the Pólya urn Gibbs sampler, the function \fct{DPMcdensity} implements the exact same model as the function \fct{DPcdensity} in the \pkg{DPpackage} \proglang{R} package, except that the latter does not provide conditional CDF estimation. We examine the run time of these two functions by running \fct{DPcdensity} and \fct{DPMcdensity} with different samplers, respectively, on Dunson's example (\ref{exp:3}). Only conditional density and conditional mean are estimated and the same grid points, hyper-parameters and MCMC settings as \code{CdDpmB} are used. Table~\ref{tab:time} shows the run time of different functions. As we can see, the \fct{DPMcdensity} with \code{method = "truncated"} is the most efficient and the \fct{DPMcdensity} with \code{method = "neal"} is much faster than its competitor \fct{DPcdensity}. In addition to the usage of the \pkg{Armadillo} \proglang{C++} library which provides efficient linear algebra calculations, the fast run time of \code{DPMcdensity(method = "neal")} with respect to \fct{DPcdensity} is due to the optimization of the algorithm for the estimated conditional density evaluation. Algorithm~\ref{alg:3} in Appendix~\ref{app:estimation} shows the algorithm that we use for the estimated conditional density evaluation. Step~\ref{step:5} in Algorithm~\ref{alg:3} evaluates the marginal and conditional density using the clusters parameters currently associated with at least one observation and saves the result, before the stick-breaking weights generation and clusters sampling. This avoids repeatedly evaluating the density evaluated at step~\ref{step:5}, thereby improving the efficiency of \fct{DPcdensity}. Hence, our package provides a fast implementation of the DPM of multivariate normals model in conditional density estimation, compared to the popular \pkg{DPpackage}. This is critical to the computational efficiency of the BART-DPM approach, because this approach involves multiple runs of the function \fct{DPMcdensity}.
\begin{table}[h!]
\centering
\begin{tabular}{lllp{2.5cm}}
\hline
Function       & \multicolumn{2}{c}{\fct{DPMcdensity}}  & \fct{DPcdensity}  \\
Gibbs Sampler  & Blocked~~           & Pólya urn~~          & Pólya urn~~  \\ \hline
Run Time       & 1.88 min~~          & 13.58 min~~          & 33.30 min~~ \\ \hline
\end{tabular}
\caption{\label{tab:time}Run time of the function \fct{DPMcdensity} in \pkg{BNPqte} and the function \fct{DPcdensity} in \pkg{DPpackage} for the example in (\ref{exp:3}). Only conditional density and mean are estimated. Other settings are the same across different functions. The estimated density is evaluated at $5100$ grid points. The number of total MCMC iterations is $20000$, out of which $5000$ posterior samples are kept.}
\end{table}
%


\section{BNP QTE Estimation with BART-DPM Approach} \label{sec:bnp}

In this section, we first present the BART-DPM approach to estimate QTEs for observational data and then demonstrate the corresponding \proglang{R} function in the \pkg{BNPqte} package with a simulated example.

\subsection{BART-DPM Approach Specification}

Consider a data framework $\{Y_i, \boldsymbol{x}_i, T_i\}_{i=1}^n$ in causal inference, where $Y_i$ is the observed outcome, $\boldsymbol{x}_i = (x_{i1}, \cdots, x_{ip})^{\top}$ is the observed confounders and $T_i$ is the treatment assigned to the $i^{\text{th}}$ subject with $T_i = 1$ for the active treatment and $T_i = 0$ for the control treatment. We assume the outcome to be continuous and allow the $p$ confounders to be of any type. Let $(Y_i(0), Y_i(1))$ be the potential outcomes for the $i^{\text{th}}$ subject under the control and active treatment, respectively. Under the consistency assumption, we have $Y_i = Y_i(T_i)$. The propensity score (PS) for the $i^{\text{th}}$ subject, denoted by $e(\boldsymbol{x}_i)$,  is defined as the probability of the subject being assigned to the active treatment group given a set of observed confounders, i.e., $P(T_i = 1 \, \vert \, \boldsymbol{x}_i)$.

To make inference in observational studies, \cite{xu2018bayesian} use three identifying assumptions, i.e., the stable unit treatment value assumption, the strong ignorability assumption and the overlap assumption, under which the treatment assignment is independent of potential outcomes given PS. As such, the CDF of the potential outcome $Y(t)$, $t \in \{0, 1\}$, is identified as
\begin{equation} \label{eq:21}
F_t(y) \, = \, \int P(Y < y \, \vert \, e(X) = e(\boldsymbol{x}), T = t) d R(e(\boldsymbol{x})),
\end{equation}
where $R$ is the distribution of PS in the population of interest and can be induced by the distribution of the confounders in the population of interest. The BART-DPM approach to model $F_t(y)$ is built based on the identification above.

The BART-DPM approach first fits a binary BART model (\ref{eq:4}) on the observations $\{T_i, \boldsymbol{x}_i\}_{i=1}^n$ to model the regression of the treatment assignment on confounders, i.e., 
\begin{equation} \label{eq:22}
T_i = 1 \, \vert \, \boldsymbol{x}_i \, \stackrel{\text{ind}}{\sim} \, \mathrm{Bernoulli}\left(H \left[ \sum\limits_{m=1}^M g(\boldsymbol{x}_i; \mathcal{T}_m, \boldsymbol{\mu}_m) \right] \right),
\end{equation}
where the link function $H$ is the CDF of the standard normal distribution for the probit BART model and the CDF of the logistic distribution for the logit BART model. $K$ posterior samples of the set of PS $\{e(\boldsymbol{x}_i) \}_{i=1}^n$ can be calculated by $e^{\{k\}}(\boldsymbol{x}_i) = H[\sum_{m=1}^M g(\boldsymbol{x}_i; \mathcal{T}_m^{\{k\}}, \boldsymbol{\mu}_m^{\{k\}})]$, $1 \leq k \leq K$, where $\{\mathcal{T}_m^{\{k\}}, \boldsymbol{\mu}_m^{\{k\}}\}_{m=1}^M$ is the $k^{\text{th}}$ posterior sample from the binary BART model. For simplicity, we denote $e^{\{k\}}(\boldsymbol{x}_i)$ by $e^{\{k\}}_i$ for $1 \leq k \leq K$ and $1 \leq i \leq n$.

Given each posterior sample $\{e^{\{k\}}_i \}_{i=1}^n$ of PS, the BART-DPM approach estimates the conditional CDF $P(Y < y \, \vert \, e(X) = e(\boldsymbol{x}), T = t)$ in (\ref{eq:21}) by fitting a DPM of bivariate normals model (\ref{eq:6}) on the observations $\{Y_i, H^{-1}(e^{\{k\}}_i)\}_{i \in \{i: T_i = t\}}$, i.e., $\{Y_i, \sum_{m=1}^M g(\boldsymbol{x}_i; \mathcal{T}_m^{\{k\}}, \boldsymbol{\mu}_m^{\{k\}}) \}_{i \in \{i: T_i = t\}}$, in treatment group $T=t$. The specific DPM model can be expressed as
\begin{align} \label{eq:23}
& (Y_i, H^{-1}(e^{\{k\}}_i))^{\top} \,\vert\,\boldsymbol{\mu}_{i},\Sigma_{i} \, \stackrel{\text{iid}}{\sim} \, \mathrm{Normal}\left((Y_i, H^{-1}(e^{\{k\}}_i))^{\top} \,\vert\,\boldsymbol{\mu}_{i},\Sigma_{i}\right), \\
& \left(\boldsymbol{\mu}_{i},\Sigma_{i}\right)\,\vert\,G  \, \stackrel{\text{iid}}{\sim} \, G, \qquad G\,\vert\,\alpha,G_{0} \, \sim \, \mathrm{DP}(\alpha G_{0}), \qquad i \in \{i : T_i = t\}. \nonumber
\end{align}
Posterior samples of Model~(\ref{eq:23}) are obtained by using either the Pólya urn Gibbs sampler discussed in Section~\ref{sec:neal} or the blocked Gibbs sampler discussed in Section~\ref{sec:blocked}. The conditional CDF $P(Y < y \, \vert \, e(X) = e(\boldsymbol{x}), T = t)$ can be estimated by either (\ref{eq:d6}) or (\ref{eq:19}) depending on which Gibbs sampler is used.

Once the estimated conditional CDF $P(Y < y \, \vert \, e(X) = e(\boldsymbol{x}), T = t)$ in (\ref{eq:21}) is obtained, the CDF $F_t(y)$ of the potential outcome $Y(t)$ can be approximated by integrating the estimated conditional CDF over the distribution $R$ of PS in the population of interest. If the true distribution of the confounders is known, we can estimate the CDF $F_t(y)$ by averaging the estimated conditional CDFs which are evaluated at the estimated PS of the confounders generated from its true distribution. If the true distribution of the confounders is unknown, the \pkg{BNPqte} package estimates the distribution $R$ by either the empirical distribution $R_n^{\{k\}}(\cdot) = \sum_{i=1}^n \delta_{e_i^{\{k\}}}(\cdot) / n$ or the distribution $R_n^{\{k\}}(\cdot) = \sum_{i=1}^n u_i \delta_{e_i^{\{k\}}}(\cdot)$ obtained from Bayesian bootstrap, where the weights $(u_1, \cdots, u_n)$ follows a $\mathrm{Dirichlet}(1,\cdots,1)$ distribution.

Quantiles of interest and QTEs can be computed from the estimated distributions of potential outcomes. Since we estimate the entire distributions of potential outcomes, we are able to estimate multiple QTEs  simultaneously. We summarize the complete approach in Algorithm \ref{alg:bnp}.

\begin{algorithm}
\caption{BART-DPM Approach to Estimate QTEs}\label{alg:bnp}
\begin{algorithmic}[1] 

\Procedure{QTE}{$\{Y_i, T_i, \boldsymbol{x}_i\}_{i=1}^n, p$}
    \State Fit a binary BART model (\ref{eq:22}) on $\{T_i, \boldsymbol{x}_i\}_{i=1}^n$ and obtain $K$ posterior samples $\{H^{-1}(e^{\{k\}}_i)\}_{i,k=1}^{n,K}$
    \State Create a set of grid points of $Y$ values: $(g_1, \cdots, g_S)$
    \If{$R$ is known}
        \State Generate $\tilde{n}$ i.i.d. $\tilde{\boldsymbol{x}}_{\tilde{i}}$ from the true distribution of the confounders
        \State Obtain the prediction $\{H^{-1}[e^{\{k\}}(\tilde{\boldsymbol{x}}_{\tilde{i}}) ]\}_{\tilde{i},k=1}^{\tilde{n},K}$ from the BART model in Step 2
    \Else
        \State Set $\tilde{n}=n$ and $\{H^{-1}[e^{\{k\}}(\tilde{\boldsymbol{x}}_{\tilde{i}}) ]\}_{\tilde{i},k=1}^{\tilde{n},K}=\{H^{-1}(e^{\{k\}}_i)\}_{i,k=1}^{n,K}$
    \EndIf
    \For{$k = 1, \cdots, K$}
        \For{$t = 0, 1$}
            \State Fit a DPM of bivariate normals (\ref{eq:23}) on $\{Y_i, H^{-1}(e^{\{k\}}_i)\}_{i \in \{i: T_i = t\}}$
            \If{Blocked Gibbs sampler is used}
                \State Use Algorithm~\ref{alg:2} to obtain $L$ posterior samples
                \State Use (\ref{eq:19}-\ref{eq:20}) to calculate $\{F^{\{kl\}}(g_s \, \vert \, H^{-1}[e^{\{k\}}(\tilde{\boldsymbol{x}}_{\tilde{i}}) ], T = t)\}_{\tilde{i},s,l=1}^{\tilde{n},S,L}$
            \EndIf
            \If{Pólya urn Gibbs sampler is used}
                \State Use Algorithm~\ref{alg:1} to obtain $L$ posterior samples
                \State Use Algorithm~\ref{alg:3} to calculate $\{F^{\{kl\}}(g_s \, \vert \, H^{-1}[e^{\{k\}}(\tilde{\boldsymbol{x}}_{\tilde{i}}) ], T = t)\}_{\tilde{i},s,l=1}^{\tilde{n},S,L}$
            \EndIf
        \EndFor
        \If{$R$ is known or Empirical distribution is used}
            \State Set $(u_1^{\{k\}}, \cdots, u_{\tilde{n}}^{\{k\}}) = (1/\tilde{n}, \cdots, 1/\tilde{n})$
        \Else
       	    \State Sample $(u_1^{\{k\}}, \cdots, u_{\tilde{n}}^{\{k\}})$ from $\mathrm{Dirichlet}(1,\cdots,1)$
        \EndIf
        \For{$l = 1, \cdots, L$}
            \State Calculate the CDF of $Y(t)$ as follows:
            \begin{equation*}
            F_t^{\{kl\}}(g_s) = \sum\limits_{i=1}^{\tilde{n}} u_i^{\{k\}} F^{\{kl\}}(g_s \, \vert \, H^{-1}[e^{\{k\}}(\tilde{\boldsymbol{x}}_{\tilde{i}}) ], T = t), \quad 1 \leq s \leq S,\, t \in \{0, 1\}
            \end{equation*}
            \State Find a grid point $g_{t,p}^{\{kl\}}$ such that $F_t^{\{kl\}}(g_{t,p}^{\{kl\}}) = p$ for $t \in \{0, 1\}$
            \State The $p^{\text{th}}$ quantile from the CDF $F_t^{\{kl\}}(\cdot)$ is $g_{t,p}^{\{kl\}}$ for the treatment group $T = t$
        \EndFor
    \EndFor
    \State The estimated CDF of $Y(t)$ on a grid point $g_s$ is $F_t(g_s) = \frac{1}{KL} \sum_{k,l=1}^{K,L} F_t^{\{kl\}}(g_s)$
    \State The estimated $p^{\text{th}}$ quantile for the treatment group $T=t$ is $\frac{1}{KL} \sum_{k,l=1}^{K,L} g_{t,p}^{\{kl\}}$
    \State The estimated $p^{\text{th}}$ QTE is $\frac{1}{KL} \sum_{k,l=1}^{K,L} (g_{1,p}^{\{kl\}} - g_{0,p}^{\{kl\}} )$
\EndProcedure

\end{algorithmic}
\end{algorithm}


\subsection{Illustrations}

The ultimate goal of the \pkg{BNPqte} package is to implement the BART-DPM approach of \cite{xu2018bayesian}. The \proglang{R} function \fct{qte} does this. The main arguments of \fct{qte} can be divided into three parts: data inputs, prediction settings and models specification.

The data is entered into the function \fct{qte} through the following three arguments.
\begin{itemize}
  \item \code{y} is a vector of continuous observed outcomes
  \item \code{x} is a vector or matrix of observed confounders of any type
  \item \code{treatment} is a vector of binary observed treatment assignments
\end{itemize}

The quantiles of interest are specified via the argument \code{probs} which is a vector of real numbers between $0$ and $1$. Similar to the two DPM functions, users can obtain credible intervals of the estimated distributions of potential outcomes through the arguments \code{compute.band} and \code{type.band}. Moreover, users can specify the confidence levels by setting different values to the argument \code{alphas}.

The BART and DPM models, as well as the distribution of PS in the population of interest, are specified by the following arguments.
\begin{itemize}
  \item \code{bart.link} is a string indicating the link used in the binary BART model (\ref{eq:22}); probit BART is used if \code{bart.link = "probit"} and logit BART is used if \code{bart.link = "logit"}
  \item \code{bart.params} is a list containing arguments in \fct{pbart} or \fct{lbart}; a default value will be used for an argument if it is not provided to \code{bart.params}
  \item \code{dpm.params} is a list containing arguments in \fct{DPMcdensity}; a default value will be used for an argument if it is not provided to \code{dpm.params}
  \item \code{Rdist} is a string indicating whether the distribution of PS is known or estimated; it is known if \code{Rdist = "known"}; it is estimated by the empirical distribution if \code{Rdist = "empirical"}; it is estimated by Bayesian bootstrap if \code{Rdist = "bootstrap"} which is the default value; 
  \item \code{xpred} is a vector or matrix giving the confounders values sampled from the true distribution of the confounders; the binary BART model makes prediction at \code{xpred} and the resulting values are used to evaluate the estimated conditional distributions in the DPM models; only used when \code{Rdist = "known"}
\end{itemize}

As seen in Algorithm~\ref{alg:bnp}, multiple DPM models need to be fit to obtain the conditional distributions of potential outcomes given the estimated PS. We provide parallel computing for fitting multiple DPM models in the function \fct{qte} by setting the argument \code{mc.cores} greater than $1$.

The object returned by the function \fct{qte} belongs to the class \code{"qte"}, which essentially is an \proglang{R} list containing predictions results and parameters used in the analysis. The functions \fct{predict} and \fct{plot} are also available for objects of class \code{"qte"}, through S3 Method.

In the followng, we illustrate the usage of the function \fct{qte} with a simulated example from \cite{xu2018bayesian} and present the prediction results for the BART-DPM approach using probit BART and logit BART respectively by generating $100$ replicated datasets and then running \fct{qte} with different \code{bart.link} on each dataset.

\begin{figure}[h!]
\centering
\includegraphics{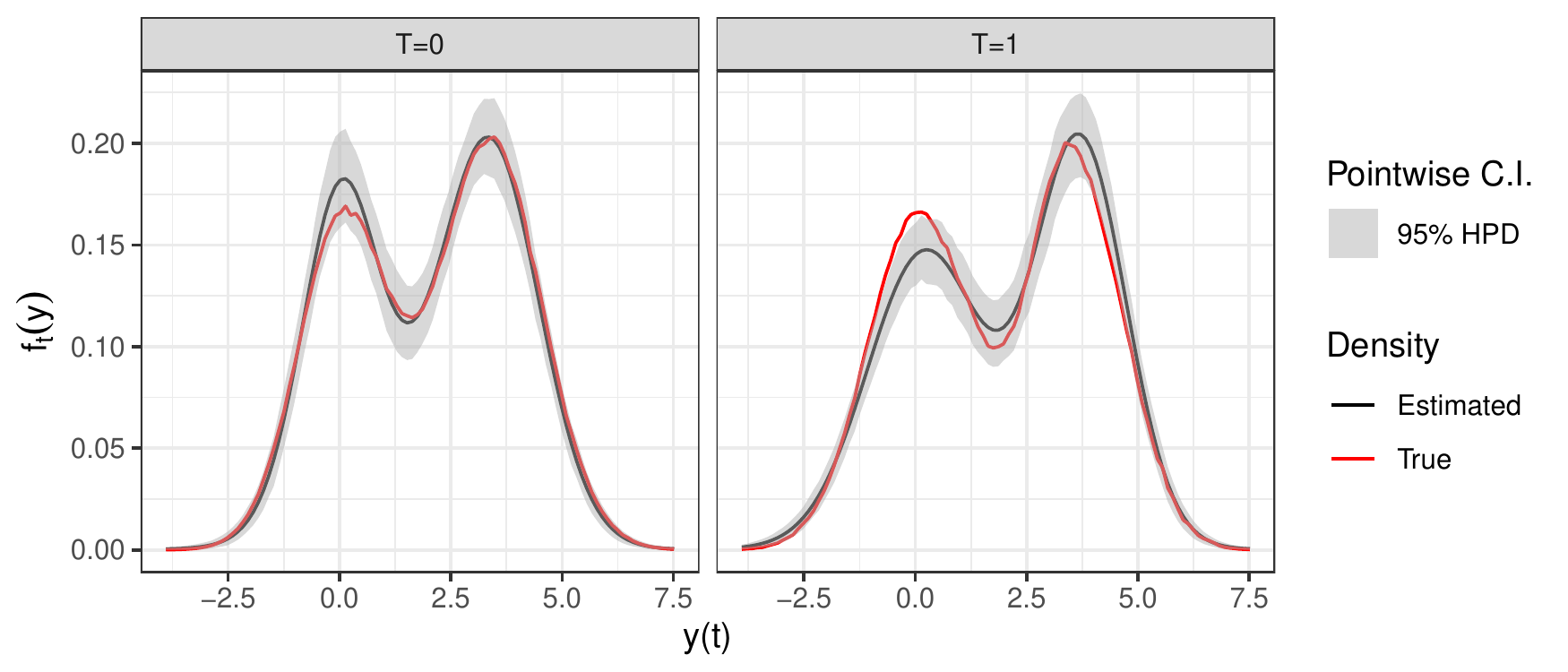}
\caption{\label{fig:qte1} Estimated density of potential outcomes for the example in (\ref{exp:4}).}
\end{figure}

\begin{figure}[h!]
\centering
\includegraphics{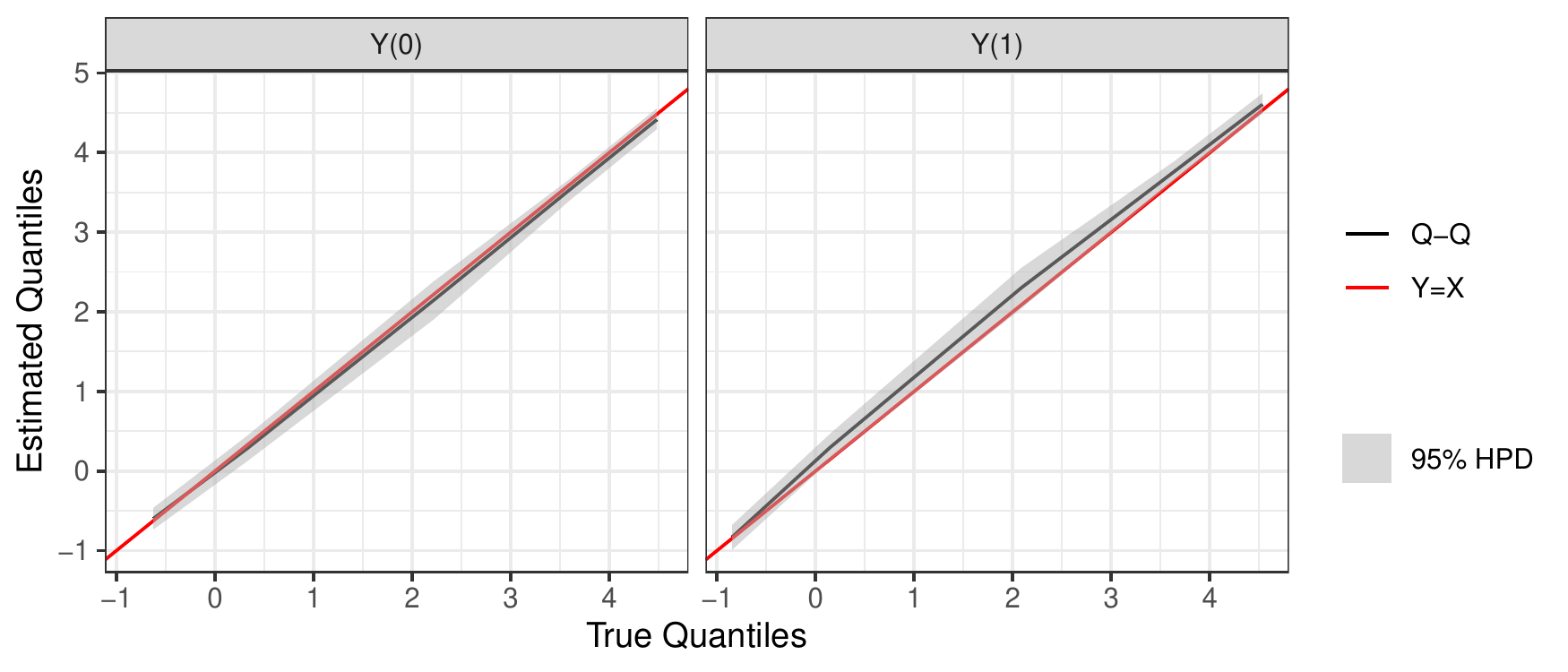}
\caption{\label{fig:qte2} Q-Q plots of potential outcomes for the example in (\ref{exp:4}).}
\end{figure}

\begin{figure}[h!]
\centering
\includegraphics{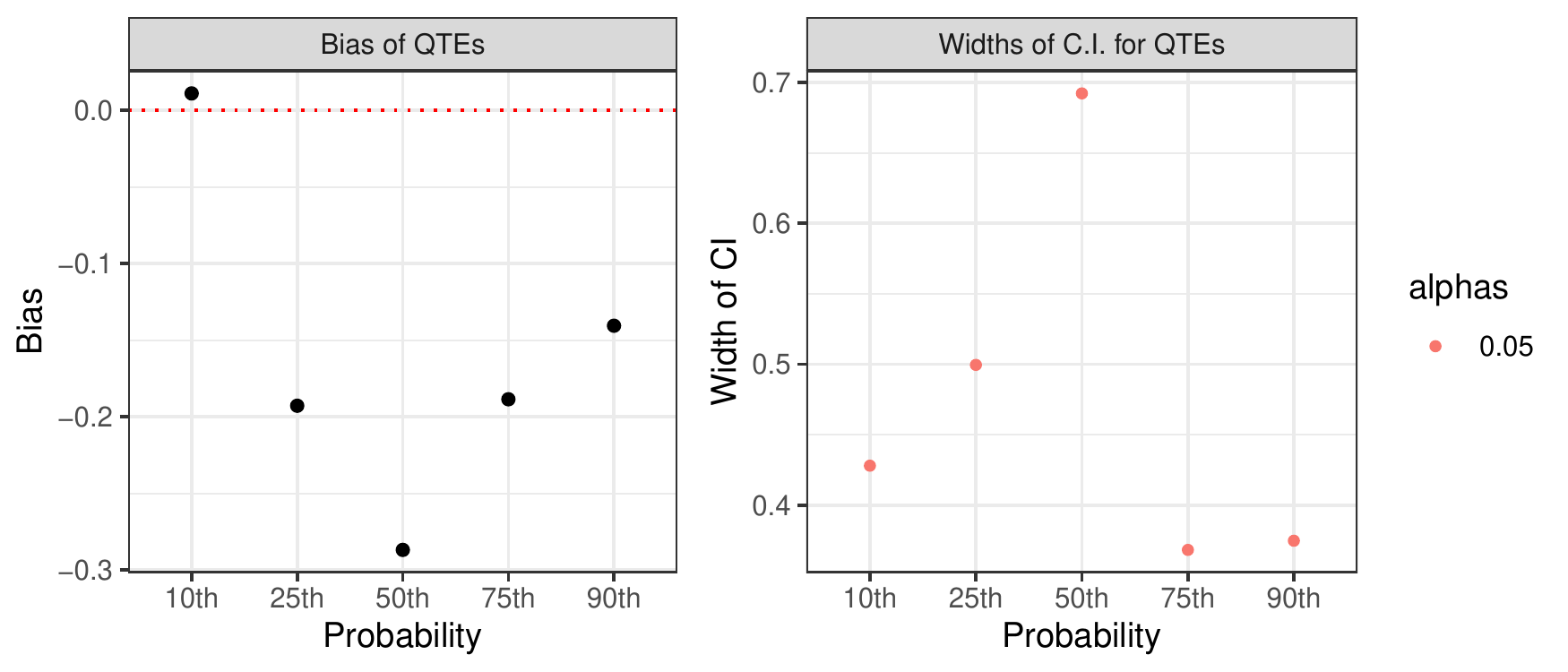}
\caption{\label{fig:qte3} Left subfigure is the bias of the averaged estimated QTEs for one replication of the example in (\ref{exp:4}) and right subfigure is the width of the $95\%$ credible intervals of the estimated QTEs for one replication of the example in (\ref{exp:4}). True $10^{\text{th}}$, $25^{\text{th}}$, $50^{\text{th}}$, $75^{\text{th}}$ and $90^{\text{th}}$ QTEs are $-0.22$, $-0.18$, $-0.13$, $0.04$ and $0.05$, respectively.}
\end{figure}

We simulate $n = 2000$ subjects with $10$ confounders $x_1, \cdots, x_{10}$ sampled from $\mathrm{Uniform}(-2, 2)$. The treatment assignment $T$ and the outcome $Y$ are generated from the following data generation process
\begin{align} \label{exp:4}
T \, \vert \, \boldsymbol{x} & \, \sim \,   \mathrm{Bernoulli} \left(0.3\sum\limits_{j=1}^{10} x_j \right) \nonumber \\
Y \, \vert \, \boldsymbol{x}, T=1 & \, \sim \,   \mathrm{expit}\left(0.5 x_3 x x_4 \right) \mathrm{Normal} \left(3 + 0.5 x_2 x_5 + 0.5 x_1^2, 0.5^2 \right) \nonumber \\
& \, \quad \, + \left[1 - \mathrm{expit} \left(0.5 x_3 x x_4 \right) \right] \mathrm{Normal} \left(-0.5 + 0.5 x_2^2 - 0.5 x_1 x_3, 0.8^2 \right) \\
Y \, \vert \, \boldsymbol{x}, T=0 & \, \sim \,  \mathrm{exp}\left(-\vert x_5 \vert \right) \mathrm{Normal}\left((\sum\limits_{j=1}^5 0.2 x_j)^4, 1 \right) \nonumber  \\
& \, \quad \, + \left(1 - \mathrm{exp}(-\vert x_5 \vert) \right) \mathrm{Normal}\left(2 + \sum\limits_{j=1}^5 0.2 x_j^2, 1 \right). \nonumber
\end{align}
This process can be done by using the function \fct{QteExample} in our package. Apart from the data (i.e., \code{$x}, \code{$treatment}, \code{$y}), this function also returns the true conditional density function of the outcome given the confounders in each treatment group: \code{$fy1} for the active treatment group and \code{fy0} for the control treatment group.
\begin{Schunk}
\begin{Sinput}
R> set.seed(0)   
R> qteData = QteExample(n = 2000)
R> names(qteData)
\end{Sinput}
\begin{Soutput}
[1] "x"         "treatment" "y0"        "y1"        "y"        
[6] "fy1"       "fy0"      
\end{Soutput}
\end{Schunk}
We run the function \fct{qte} on the data (\code{qteData$y}, \code{qteData$x}, \code{qteData$treatment}) to estimate the $10^{\text{th}}$, $25^{\text{th}}$, $50^{\text{th}}$, $75^{\text{th}}$ and $90^{\text{th}}$ QTEs (i.e., \code{probs = c(0.1, 0.25, 0.50, 0.75, 0.90)}) simultaneously. Specifically, we use the probit BART model (i.e., \code{bart.link = "probit"}) to estimate the PS. We use $50$ trees (i.e., \code{bart.params$ntree = 50}) in the BART model and run the MCMC for $1000$ iterations and keep every $100^{\text{th}}$ posterior sample (i.e., \code{bart.params$keepevery = 100}) after the first $500$ iterations (i.e., \code{bart.params$nskip = 500}). Other hyper-parameters of BART are set to default values. For the DPM model, we use the blocked Gibbs sampler (i.e., \code{dpm.params$method = "truncated"}) with $50$ clusters (i.e., \code{dpm.params$nclusters = 50}) and run $900$ MCMC iterations with the first $500$ samples burned in (i.e., \code{dpm.params$nskip = 500}) and every $2^{\text{rd}}$ sample saved (i.e., \code{dpm.params$keepevery = 2}). Other hyper-parameters of the DPM model are set as suggested in Appendix~\ref{app:hyper}. Besides the estimated quantiles, we require the function to return the estimated density and CDF of potential outcomes by setting \code{dpm.params$type.pred = c("cdf", "pdf")} and their $95\%$ pointwise credible intervals are also returned via the setting \code{compute.band = TRUE, type.band = "HPD", alphas = c(0.05)}. We use Bayesian bootstrap (i.e., \code{Rdist = "bootstrap"}) to estimate the distribution of PS in the population of interest. The following code realizes the settings described above.
\begin{Schunk}
\begin{Sinput}
R> qteFit = qte(y = qteData$y, x = qteData$x, treatment = qteData$treatment,
+               probs = c(0.1, 0.25, 0.50, 0.75, 0.90), 
+               compute.band = TRUE, type.band = "HPD", alphas = c(0.05), 
+               bart.link = "probit", 
+               bart.params = list(ntrees = 50, nskip = 500, ndpost = 5,
+                                  keepevery = 100), 
+               dpm.params = list(method = "truncated", nclusters = 50, 
+                                 type.pred = c("cdf", "pdf"), ngrid = 100,
+                                 nskip = 500, ndpost = 200, keepevery = 2), 
+               Rdist = "bootstrap", mc.cores = 1)   
\end{Sinput}
\end{Schunk}
The returned object \code{qteFit} is of class \code{"qte"} and has the following components.
\begin{Schunk}
\begin{Sinput}
R> names(qteFit)   
\end{Sinput}
\begin{Soutput}
 [1] "propensity"              "bart.parmas"            
 [3] "grid"                    "xpred"                  
 [5] "type.pred"               "compute.band"           
 [7] "type.band"               "alphas"                 
 [9] "probs"                   "control.cdfs"           
[11] "treatment.cdfs"          "control.quantiles.avg"  
[13] "treatment.quantiles.avg" "qtes.avg"               
[15] "control.quantiles.ci"    "treatment.quantiles.ci" 
[17] "qtes.ci"                 "control.pdfs.avg"       
[19] "treatment.pdfs.avg"      "control.pdfs.ci"        
[21] "treatment.pdfs.ci"       "n0"                     
[23] "n1"                      "p"                      
[25] "dpm.params"             
\end{Soutput}
\end{Schunk}
Let us look at a couple of key components.
\begin{itemize}
  \item \code{$qtes.avg} is a vector of the averaged QTEs and the $i^{\text{th}}$ \code{$qtes.avg} corresponds to the $i^{\text{th}}$ \code{probs}
  \item \code{$qtes.ci} is a matrix representing the credible intervals of estimated QTEs; the $i^{\text{th}}$ row corresponds to the $i^{\text{th}}$ \code{probs}; the first (second) column refers to the lower (upper) bound; other returns with ".ci" suffix have similar definitions
  \item \code{$control.pdfs.avg} (\code{$treatment.pdfs.avg}) is a vector of the averaged estimated density of the potential outcome under the control (active) treatment group, which is evaluated at \code{$grid}
  \item \code{$control.cdfs} (\code{$treatment.cdfs}) is a matrix of the estimated CDF of the potential outcome under the control (active) treatment group, which is evaluated at \code{$grid}; each row is for a posterior sample and each column is for a grid point
\end{itemize}
The prediction results of \fct{qte} can be visualized by simply applying the \fct{plot} function to the object \code{qteFit}, i.e., \code{plot(qteFit)}. The estimated density of potential outcomes (see Figure~\ref{fig:qte1}) and the width of the credible intervals of the estimated QTEs (see Figure~\ref{fig:qte3}) are plotted by \fct{plot}. If the true quantiles of interest are provided, the \fct{plot} function also returns a Q-Q plot (see Figure~\ref{fig:qte2}) and a plot for the bias of the estimated QTEs (see Figure~\ref{fig:qte3}). The following code shows the usage of the \fct{plot} function for objects of class \code{"qte"}. Since the true density and the true quantiles of the potential outcomes for the example in (\ref{exp:4}) are not analytically achievable, we use the functions \code{qteData$fy0} and \code{qteData$fy0} to simulate the true density at grid points and generate the true quantiles from a large sample data.
\begin{Schunk}
\begin{Sinput}
R> truefy0 = truefy1 = c()
R> for (i in 1 : 100) {
+    truefy0[i] = qteData$fy0(qteFit$grid[i], 10000)
+    truefy1[i] = qteData$fy1(qteFit$grid[i], 10000)
+  }
R> Data0 = SampleFun5(n = 10000000)
R> q0 = quantile(x = Data0$y0, probs = c(0.1, 0.25, 0.5, 0.75, 0.9))
R> q1 = quantile(x = Data0$y1, probs = c(0.1, 0.25, 0.5, 0.75, 0.9))
R> plot(qteFit, true.fy1 = truefy1, true.fy0 = truefy0,
+       true.quantile1 = q1, true.quantile0 = q0, true.qte = q1 - q0)
\end{Sinput}
\end{Schunk}
We also provide the S3 \fct{predict} function for objects of class \code{"qte"}. Given the fitted object from the \fct{qte} function, users can estimate more quantiles as well as QTEs from the estimated CDFs of potential outcomes. Taking the object \code{qteFit} as an example, we show the syntax of the \fct{predict} function for objects of class \code{"qte"} as follows.
\begin{Schunk}
\begin{Sinput}
R> predict(object = qteFit, probs = seq(0.1, 0.99, 100))
\end{Sinput}
\end{Schunk}
We conclude this section with simulation results based on $100$ repeatedly simulated datasets for the BART-DPM approach using probit BART and logit BART respectively. Specifically, we generate $100$ datasets according to the data generation process in (\ref{exp:4}) and run the \fct{qte} function on each dataset with probit BART, i.e., \code{bart.link = "probit}, and logit BART, i.e., \code{bart.link = "logit"}, respectively. Arguments except \code{bart.link} are set the same as the example above. Table~\ref{tab:qte} shows the result. We can see that the two models yield similar estimation for QTEs.
%
%
\begin{table}[h!]
\centering
\begin{tabular}{lllp{3.2cm}}
\hline
Quantiles       & Truth     & Probit BART                   & Logit BART  \\ \hline
10th            & $-0.22$   & $-0.22(-0.44, 0.01)$          & $-0.22(-0.44, 0.01)$ \\ 
25th            & $-0.18$   & $-0.18(-0.44, 0.08)$          & $-0.18(-0.44, 0.09)$ \\ 
50th            & $-0.13$   & $-0.11(-0.51, 0.29)$          & $-0.12(-0.52, 0.28)$ \\ 
75th            & $0.04$    & $0.06(-0.16, 0.28)$           & $0.05(-0.17, 0.26)$ \\ 
90th            & $0.05$    & $0.06(-0.16, 0.28)$           & $0.05(-0.16, 0.27)$ \\ \hline
\end{tabular}
\caption{\label{tab:qte} Comparison of the average QTEs and $95\%$ CIs across $100$ replications between the BART-DPM approaches using probit BART and logit BART for the example in (\ref{exp:4}).}
\end{table}
%


\section{Convergence Diagnostics for DPM Models} \label{sec:diagnostics}

In this section, we empirically compare the convergence of the MCMC chains between the Pólya urn Gibbs sampler (Algorithm~\ref{alg:1}) and the blocked Gibbs sampler (Algorithm~\ref{alg:2}). The cluster-specific parameters such as $\boldsymbol{\zeta}_k$, $\Omega_k$ and $\omega_k$ cannot be directly used for assessing the convergence due to label switching. Thus, it is more appropriate to examine a quantity that is not affected by the issue, such as a global (hyper-)parameter. In the following analysis, we investigate the autocorrelation of the global parameter $\alpha$, the global hyper-parameters $\lambda$, $\boldsymbol{m}$ and $\Psi$ and the logarithm likelihood, to assess the mixing of the MCMC chains generated by different Gibbs samplers. The posterior samples of the global (hyper-)parameters can be found in the sub-list \code{$posterior} of an object returned by the \fct{DPMdensity} or \fct{DPMcdensity} function and the posterior samples of the logarithm likelihood are calculated and returned to the sub-list \code{$posterior} when the argument \code{diag} of the function \fct{DPMdensity} or \fct{DPMcdensity} is set to \code{TRUE}. When \code{diag = TRUE} and \code{method = "truncated"}, the posterior samples of the logarithm marginal partition posterior $\log(f(\{\kappa_i\}_{i=1}^n \, \vert \, \{\boldsymbol{z}_i\}_{i=1}^n))$ suggested by \cite{hastie2015sampling} are also returned to the sub-list \code{$posterior}. This quantity represents the posterior distribution of the allocations, with all the other parameters integrated out. However, the marginal partition posterior does not have a closed form for the limit model of Model~(\ref{eq:9}), so we do not provide it for the Pólya urn Gibbs sampler, nor investigate it here. We show the calculation of the marginal partition posterior as well as the likelihood in Appendix~\ref{app:diag}. For more convergence diagnostics, we recommend using the \pkg{coda} \proglang{R} package (\cite{coda}) which provides functions for summarizing and plotting the outputs from a MCMC chain and diagnostics tests of convergence for a MCMC chain.

\begin{figure}[h!]
\centering
\includegraphics{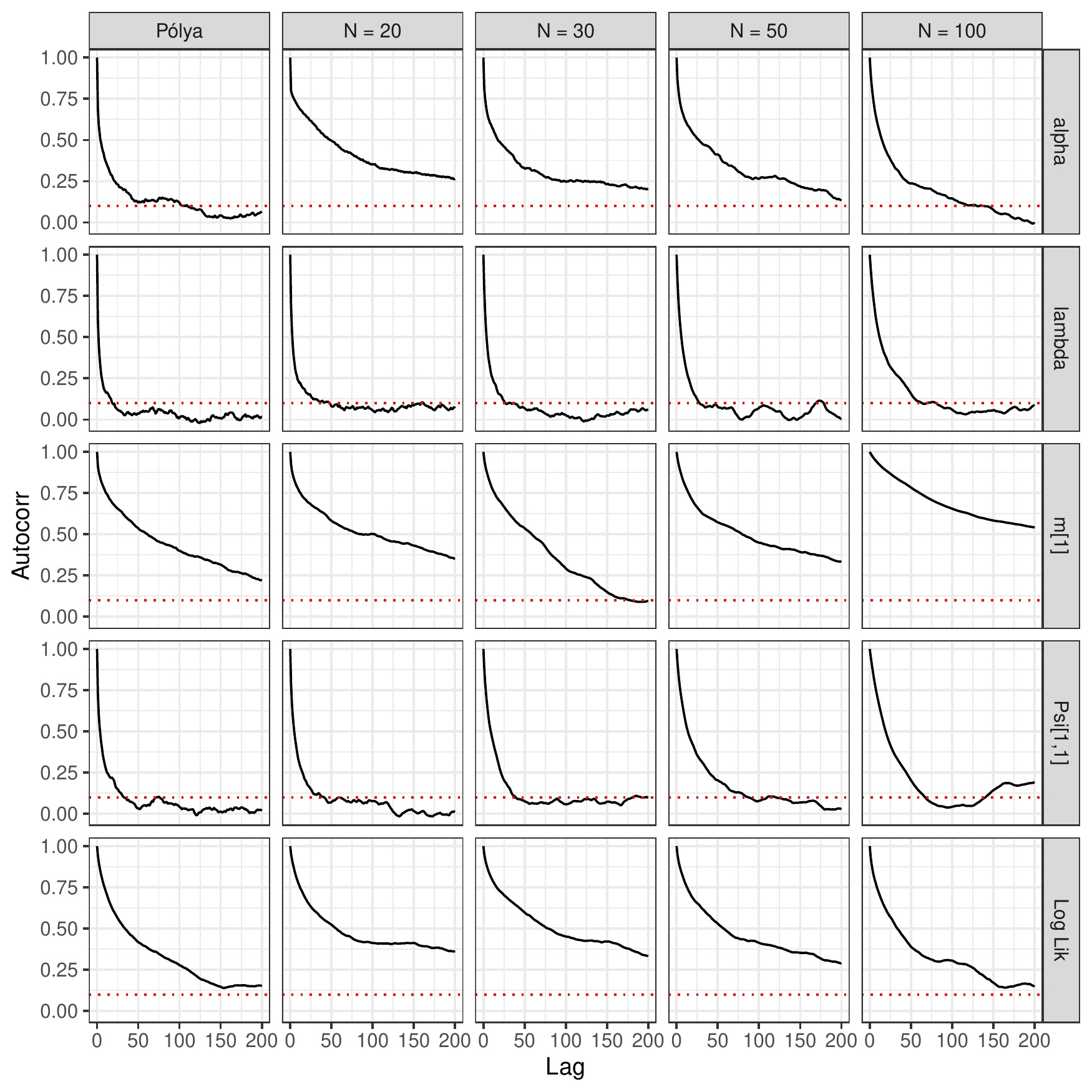}
\caption{\label{fig:auto} Autocorrelation for different parameters (the parameter $\alpha$, the global hyper-parameters $\lambda$, $\boldsymbol{m}$ and $\Psi$ and the logarithm likelihood) and different samplers (the Pólya urn Gibbs sampler and the blocked Gibbs sampler with $20$, $30$, $50$ and $100$ clusters). Rows are for parameters and columns are for samplers.}
\end{figure}

Now, we revist the simulation example in the joint density estimation, i.e., the mixture of three bivariate normals. We run the function \fct{DPMdensity} with the Pólya urn Gibb sampler and the blocked Gibbs sampler respectively on the data \code{DpmData1$y} generated from (\ref{exp:2}). For the blocked Gibbs sampler, we run the function four times with different number of clusters: $N \in \{20, 30, 50, 100\}$. For each MCMC chain, we skip the first $10000$ iterations and save the subsequent $10000$ posterior samples.
\begin{Schunk}
\begin{Sinput}
R> BFit = list()
R> Nvec = c(20, 30, 50, 100)
R> for (i in 1 : 4) {
+    BFit[[i]] = DPMdensity(DpmData1$y, ngrid = 0, method = "truncated",
+                           nclusters = Nvec[i],  diag = TRUE,
+                           nskip = 10000, ndpost = 10000, keepevery = 1)
+  }
R> PFit = DPMdensity(DpmData1$y, ngrid = 0, method = "neal", diag = TRUE,
+                    nskip = 10000, ndpost = 10000, keepevery = 1)
\end{Sinput}
\end{Schunk}
Taking the object \code{PFit} as an example, the parameters of interest can be obtained by the following code.
\begin{Schunk}
\begin{Sinput}
R> alpha = PFit$posterior$alpha
R> lambda = PFit$posterior$lambda
R> m = PFit$posterior$m
R> Psi = PFit$posterior$Psi
R> llik = PFit$posterior$ylogliks
R> length(alpha)
\end{Sinput}
\begin{Soutput}
[1] 10000
\end{Soutput}
\end{Schunk}
Figure~\ref{fig:auto} shows the autocorrelation for different parameters and different samplers. In general, the MCMC chains generated by the Pólya urn Gibb sampler mix better than those generated by the blocked Gibbs sampler. From the first row of Figure~\ref{fig:auto}, we can see that as $N$ increases, the autocorrelation of $\alpha$ sampled by the blocked Gibbs sampler decreases fast and becomes similar to that sampled by the Pólya urn Gibbs sampler. This is because as $N$ increases, the mixing weights (\ref{eq:13}) used to update $\alpha$ converge to the stick-breaking weights of DP (\ref{eq:11}) and therefore the distribution used to update $\alpha$ in the blocked Gibb sampling becomes similar to that in the Pólya urn Gibbs sampling. From the second to the fourth row of Figure~\ref{fig:auto}, we find that the chains of hyper-parameters based on the blocked Gibbs sampling mix better for a smaller $N$. Let $\eta$ denote a hyper-parameter of the base distribution $G_0$. Both of the two samplers update $\eta$ based on the product distribution of the hyper-prior on $\eta$ and the base distribution $G_0$, except that the Pólya urn Gibbs sampler only includes $G_0$ of the parameters associated with some observations while the blocked Gibbs sampler includes $G_0$ of all the parameters. The parameters included by the blocked Gibbs sampler but not related to any observation are sampled from their prior distributions and shared by different $\eta$'s across the MCMC iterations, thereby making the chain of the hyper-parameter mix slowly. As $N$ decreases, there are fewer empty clusters, so the chain mixes better. The fifth row of Figure~\ref{fig:auto} shows the autocorrelation of logarithm likelihood. As we can see, when $N$ gets greater, the autocorrelation of the log likelihood of the blocked Gibbs sampler becomes similar to that of the Pólya urn Gibbs sampler, which can be explained by the almost sure convergence of the truncated DPM model.
%


\section{Summary} \label{sec:summary}

This article reviews the models and algorithms implemented in the \pkg{BNPqte} \proglang{R} package, introduces the usage of the \proglang{R} functions with simulated examples, and assesses the mixing properties of the two Gibbs sampling algorithms used for the DPM models. As software for causal inference, \pkg{BNPqte} implements a Bayesian approach to estimate QTEs and provides \proglang{R} users who are interested in QTE estimation with an alternative \proglang{R} package to those \proglang{R} packages implementing frequentist methods. Moreover, \pkg{BNPqte} provides a fast implementation of the DPM of multivariate normals model in density estimation, especially in conditional destimation, compared to the popular \pkg{DPpackage} \proglang{R} package. Finally, \pkg{BNPqte} modifies the functions from the \pkg{BART} \proglang{R} package to improve them for data with mixed-type predictors and to obtain the optimal posterior contraction rate.


\section*{Computational Details}

The results in this article were obtained using
\proglang{R}~3.6.0 with the
\pkg{BNPqte} \proglang{R} package. \proglang{R} itself and all packages used, except \pkg{BNPqte} and \pkg{DPpackage}, are available from the Comprehensive
\proglang{R} Archive Network (CRAN) at
\url{https://CRAN.R-project.org/}. \pkg{BNPqte} is available at \url{https://github.com/chujiluo/BNPqte}. \pkg{DPpackage} is archived form CRAN, but the source code is available at \url{https://github.com/cran/DPpackage}.


\bibliography{refs}



\newpage

\begin{appendix}


\section{Hyper-Parameters Specification for DPM Models} \label{app:hyper}

We adopt the specification of \cite{taddy2008bayesian} to choose hyper-parameters, which ensures that the hyper-priors are appropriately diffuse, are scaled to the data, and provide minimal prior information.

Following the notation of Section~\ref{sec:dpmm}, let $c_y$ and $c_{x_j}$ denote the mean of $\{Y_i\}_{i=1}^n$ and $\{x_{ij}\}_{i=1}^n$, $j=1,\cdots,p$, and also $r_y$ and $r_{x_j}$ be the corresponding range, i.e., difference between the max and min values. Let $\boldsymbol{c} = (c_y, c_{x_1}, \cdots, c_{x_p})$ and denote by $R$ the $d \times d$ diagonal matrix with diagonal elements $\{(r_y/4)^2, (r_{x_1}/4)^2, \cdots, (r_{x_p}/4)^2\}$, which represent prior guesses for the mean and variability of the data, respectively.

We first specify the hyper-parameters for the DPM model without hyper-priors, i.e., Model~(\ref{eq:6}-\ref{eq:7}). Based on the form of $G_0$, we can obtain the marginal prior moments for $\boldsymbol{\mu}$, i.e., $\E(\boldsymbol{\mu})=\boldsymbol{m}$ and $\mathrm{Cov}(\boldsymbol{\mu}) = \frac{\Psi}{\lambda (\nu-d-1)}$, as well as the marginal mean for $\Sigma$, i.e., $\E(\Sigma) = \frac{\Psi}{\nu-d-1}$. We take $\E(\boldsymbol{\mu}) = \boldsymbol{c}$, set $\mathrm{Cov}(\boldsymbol{\mu}) = 2R$ using a variance inflation factor of $2$, specify $\E(\Sigma) = R$, and finally choose a value of $\nu$ which yields a more dispersed prior for $\Sigma$. Hence, by default, we set the hyper-parameters of Model~(\ref{eq:6}-\ref{eq:7}) as follows:
$$\boldsymbol{m} = \boldsymbol{c}, \quad \lambda = 0.5, \quad \nu = d+2, \quad \Psi = R.$$

For the hierarchical model~(\ref{eq:6}-\ref{eq:8}), the marginal prior moments for $\boldsymbol{\mu}$ and $\Sigma$ are slightly changed by the hyper-priors in (\ref{eq:8}), i.e., $\E(\boldsymbol{\mu})=\boldsymbol{m}_0$, $\mathrm{Cov}(\boldsymbol{\mu}) = S_0 + \frac{\gamma_2}{\gamma_1 - 1}\times \frac{\nu_0 \Psi_0}{\nu-d-1}$ and $\E(\Sigma) = \frac{\nu_0 \Psi_0}{\nu-d-1}$. With the same idea to estimate $\E(\boldsymbol{\mu})$ by $\boldsymbol{c}$, $\mathrm{Cov}(\boldsymbol{\mu})$ by $2R$ and $\E(\Sigma)$ by $R$, we obtain $\boldsymbol{m}_0 = \boldsymbol{c}$, $S_0 + \frac{\gamma_2}{\gamma_1 - 1}\times \frac{\nu_0 \Psi_0}{\nu-d-1} = 2R$ and $\frac{\nu_0 \Psi_0}{\nu-d-1} = R$. By setting $S_0 = R$, we have $\frac{\gamma_2}{\gamma_1 - 1} = 1$ and $\frac{\nu_0 \Psi_0}{\nu-d-1} = R$. By default, we let $\gamma_1 = 3$ and $\gamma_2 = 2$ to satisfy $\frac{\gamma_2}{\gamma_1 - 1} = 1$. We also set $\nu = \nu_0 = d+2$ to allow the largest dispersion. In conclusion, by default, we specify the hyper-parameters for Model~(\ref{eq:6}-\ref{eq:8}) as follows:
$$\boldsymbol{m}_0 = \boldsymbol{c}, \quad S_0 = R, \quad \gamma_1 = 3, \quad \gamma_2 = 2, \quad  \nu = \nu_0 = d+2, \quad \Psi_0 = R/\nu_0.$$

Regarding the choice of the concentration parameter $\alpha$, we set $\alpha=10$ for the Model~(\ref{eq:6}-\ref{eq:7}) and $a_0=10, b_0=1$ for the hierarchical model~(\ref{eq:6}-\ref{eq:8}), to allow higher prior probability for larger number of distinct clusters (\cite{antoniak1974mixtures}).

\section{Posterior Computation for Pólya Urn Gibbs Sampling} \label{app:neal}

In this section, we provide the details for sampling the posterior distribution of $\phi_k = \left(\boldsymbol{\zeta}_k, \Omega_k \right)$'s, $\alpha$ and other hyperparameters for the DPM of multivariate normals model (Model~(\ref{eq:6}-\ref{eq:7}) or Model~(\ref{eq:6}-\ref{eq:8})), using Algorithm \ref{alg:1}, i.e., Algorithm 8 of \cite{neal2000markov}.

It is not possible to explicitly represent the infinite number of $\left(\boldsymbol{\zeta}_k, \Omega_k \right)$'s for the DPM model, so instead, Algorithm \ref{alg:1} updates only those $\left(\boldsymbol{\zeta}_k, \Omega_k \right)$'s currently associated with some observations. Based on (\ref{eq:6}) and (\ref{eq:7}), we sample $\left(\boldsymbol{\zeta}_k, \Omega_k \right)$ ($k \in \{\kappa_1, \cdots, \kappa_n\}$) from their conjugate poterior distribution:
\begin{align} \label{eq:param}
f(\boldsymbol{\zeta}_{k},\Omega_{k}\,\vert\,\{\boldsymbol{z}_{i}:\kappa_{i}=k\}) & \propto f_{\mathrm{Normal}}(\boldsymbol{\zeta}_{k}\,\vert\,\boldsymbol{m},\frac{1}{\lambda}\Omega_{k})\times f_{\mathrm{IW}}(\Omega_{k}\,\vert\,\nu,\Psi)\times\prod\limits_{\{i:\kappa_{i}=k\}}f_{\mathrm{Normal}}(\boldsymbol{z}_{i}\,\vert\,\boldsymbol{\zeta}_{k},\Omega_{k}) \nonumber \\
 & \propto f_{\mathrm{Normal}}(\boldsymbol{\zeta}_{k}\,\vert\,\boldsymbol{m}^{*},\frac{1}{\lambda^{*}}\Omega_{k})\times f_{\mathrm{IW}}(\Omega_{k}\,\vert\,\nu^{*},\Psi^{*}),
\end{align}
where
\begin{align} \label{eq:param1}
\boldsymbol{m}^{*} & =\frac{\lambda\boldsymbol{m}+n_k \bar{z}_k}{\lambda+n_{k}}, \qquad \lambda^{*} =\lambda+n_{k}, \qquad \nu^{*} =\nu+n_{k}, \nonumber \\
\Psi^{*} & =\Psi+\sum\limits _{\{i:\kappa_{i}=k\}}(\boldsymbol{z}_{i}-\bar{z}_{k})(\boldsymbol{z}_{i}-\bar{z}_{k})^{\top}+\frac{\lambda n_{k}}{\lambda+n_k}(\bar{z}_{k}-\boldsymbol{m})(\bar{z}_{k}-\boldsymbol{m})^{\top},\\
n_{k} & =\vert\{i:\kappa_{i}=k\}\vert , \qquad \bar{z}_k = \frac{1}{n_k} \sum\limits_{\{i:\kappa_i = k\}} \boldsymbol{z}_i. \nonumber
\end{align}
Note that $f$ without a subscript in (\ref{eq:param}) represents a general density and that $f$ with a subscript represents a specific density specified by the subscript. Parameters after $\vert$ are parameterized as usual. For example, $f_{\mathrm{Normal}}(\boldsymbol{z} \, \vert \, \boldsymbol{\zeta}, \Omega)$ means a normal density with mean equal to $\boldsymbol{\zeta}$ and covariance matrix equal to $\Omega$ for the observation $\boldsymbol{z}$. We follow this notation in this section and subsequent sections.

If the hierarchical structure (\ref{eq:8}) is assumed for the DPM model, we update $\alpha$, $\boldsymbol{m}$, $\lambda$ and $\Psi$ sequentially at each iteration of MCMC as follows.
\begin{enumerate}
  \item The concentration parameter $\alpha$ is updated in two steps, using the approach of \cite{escobar1995bayesian}:
  \begin{enumerate}
    \item sample $\eta$ from $\mathrm{Beta}(\alpha + 1, n)$ and calculate a weight $\pi_{\eta}$ such that $\frac{\pi_{\eta}}{1 - \pi_{\eta}} = \frac{a_0 + K^* - 1}{n(b_0 - \log(\eta))}$, where $K^*$ is the number of distinct values in $\{\kappa_i\}_{i=1}^n$;
    \item sample a new value for $\alpha$ from a mixture of gamma distribution:
    $$\pi_{\eta} \mathrm{Gamma}(a_0 + K^*, b_0 - \log(\eta))  + (1 - \pi_{\eta}) \mathrm{Gamma}(a_0 + K^* - 1, b_0 - \log(\eta)).$$
  \end{enumerate}
  \item Sample a new value for the mean vector $\boldsymbol{m}$ of the base distribution $G_0$ from
  \begin{align} \label{eq:m}
  f(\boldsymbol{m} \, \vert \, \{\boldsymbol{\zeta}_k, \Omega_k\}_{k \in \{\kappa_i\}_{i=1}^n},  & \{\kappa_i\}_{i=1}^n, \lambda, \boldsymbol{m_0}, S_0) \nonumber \\
  & \propto f_{\mathrm{Normal}}(\boldsymbol{m} \, \vert \, \boldsymbol{m_0}, S_0) \times \left[ \prod\limits_{k \in \{\kappa_i\}_{i=1}^n} f_{\mathrm{Normal}} (\boldsymbol{\zeta}_k \, \vert \, \boldsymbol{m}, \frac{1}{\lambda} \Omega_k) \right] \nonumber \\
  & \propto f_{\mathrm{Normal}} (\boldsymbol{m} \, \vert \, \boldsymbol{m_0^*}, S_0^*)
  \end{align}
  where $(S_0^*)^{-1} = \lambda  \left( \sum\limits_{k \in \{\kappa_i\}_{i=1}^n} \Omega_k^{-1} \right) + S_0^{-1}$ and $\boldsymbol{m_0^*} = S_0^* \left[ \lambda \left( \sum\limits_{k \in \{\kappa_i\}_{i=1}^n} \Omega_k^{-1}\boldsymbol{\zeta}_k \right) + S_0^{-1}\boldsymbol{m_0} \right]$.
  \item Sample a new value for the scale parameter $\lambda$ of the base distribution $G_0$ from
  \begin{align} \label{eq:lambda}
  f(\lambda \, \vert \, \{\boldsymbol{\zeta}_k, \Omega_k\}_{k \in \{\kappa_i\}_{i=1}^n},  & \{\kappa_i\}_{i=1}^n, \boldsymbol{m}, \gamma_1, \gamma_2) \nonumber \\
  & \propto f_{\mathrm{Gamma}}(\lambda \, \vert \, \gamma_1, \gamma_2) \times \prod\limits_{k \in \{\kappa_i\}_{i=1}^n} f_{\mathrm{Normal}} (\boldsymbol{\zeta}_k \, \vert \, \boldsymbol{m}, \frac{1}{\lambda} \Omega_k) \nonumber \\
  & \propto f_{\mathrm{Gamma}}(\lambda \, \vert \, \gamma_1^*, \gamma_2^*)
  \end{align}
  where $\gamma_1^* = \gamma_1 + \frac{dK^*}{2}$ and $\gamma_2^* = \gamma_2 + \frac{1}{2} \sum\limits_{k \in \{\kappa_i\}_{i=1}^n} (\boldsymbol{\zeta}_k - \boldsymbol{m})^{\top} \Omega_k^{-1} (\boldsymbol{\zeta}_k - \boldsymbol{m})$.
  \item Sample a new value for the scale matrix $\Psi$ of the base distribution $G_0$ from
  \begin{align} \label{eq:psi}
  f(\Psi \, \vert \, \{\Omega_k\}_{k \in \{\kappa_i\}_{i=1}^n}, & \{\kappa_i\}_{i=1}^n, \nu, \nu_0, \Psi_0) \nonumber \\
  & \propto f_{\mathrm{Wishart}}(\Psi \, \vert \,  \nu_0, \Psi_0) \times \prod\limits_{k \in \{\kappa_i\}_{i=1}^n} f_{\mathrm{IW}} (\Omega_k \, \vert \, \nu, \Psi) \nonumber \\
  & \propto f_{\mathrm{Wishart}}(\Psi \, \vert \,  \nu_0^*, \Psi_0^*)
  \end{align}
  where $\nu_0^* = \nu K^* + \nu_0$ and $\Psi_0^* = \left(\Phi_0^{-1} +  \sum\limits_{k \in \{\kappa_i\}_{i=1}^n} \Omega_k^{-1} \right)^{-1}$.
\end{enumerate}

Note that the auxiliary parameters created during the update of $\kappa_i$'s are neither updated nor involved in the update of hyper-parameters, which facilitates fast convergence of hyper-parameters.


\section{Posterior Computation for Blocked Gibbs Sampling} \label{app:truncated}

In this section, we provide details for posterior sampling for model~(\ref{eq:14}) with $G_0$ assumed to be (\ref{eq:7}) and the hyper-parameters of $G_0$ and the parameter $\alpha$ assumed to be (\ref{eq:8}), using Algorithm \ref{alg:2}, i.e., the blocked Gibbs sampler of \cite{ishwaran2001gibbs}. The posterior samples from this sampler are used for approximating posterior samples of the corresponding DPM of multivariate normals model.

Similar to (\ref{eq:15}), the full posterior distribution of model~(\ref{eq:14}, \ref{eq:7}, \ref{eq:8}) can be expressed as follows
\begin{align*}
f(\{\phi_k & \}_{k=1}^N, \{\kappa_i\}_{i=1}^n, \boldsymbol{\omega}, \alpha, \boldsymbol{m}, \lambda, \Psi \, \vert \, \{\boldsymbol{z}_i\}_{i=1}^n, a_0, b_0, \boldsymbol{m}_0, S_0, \gamma_1, \gamma_2, \nu, \nu_0, \Psi_0) \, \\
\propto \,  & \left[ \prod\limits_{j=1}^{K^*} f(\phi_{\kappa_j^*} \, \vert \, \boldsymbol{m}, \lambda, \nu, \Psi) \prod\limits_{\{i:\kappa_i = \kappa_j^*\}} f(\boldsymbol{z}_i \, \vert \, \phi_{\kappa_j^*}) \right] \times \left[ \prod\limits_{k \notin \boldsymbol{\kappa}^*} f(\phi_k \, \vert \, \boldsymbol{m}, \lambda, \nu, \Psi) \right]\\
&  \times \left[ \prod\limits_{i=1}^n f(\kappa_i \, \vert \, \boldsymbol{\omega}) \right] \times f(\boldsymbol{\omega} \, \vert \, \alpha) \times f(\alpha \, \vert \, a_0, b_0) \times f(\boldsymbol{m} \, \vert \, \boldsymbol{m}_0, S_0) \times f(\lambda \, \vert \, \gamma_1, \gamma_2) \times f(\Psi \, \vert \, \nu_0, \Psi_0),
\end{align*}
where $\phi_k = (\boldsymbol{\zeta}_k, \Omega_k)$ for $1 \leq k \leq N$. Based on the formulation above, we update $\{\phi_k \}_{k=1}^N$, $\{\kappa_i\}_{i=1}^n, \boldsymbol{\omega}, \alpha, \boldsymbol{m}, \lambda, \Psi$ sequentially at each iteration of MCMC as follows.
\begin{enumerate}
  \item Sample a new value for each $(\boldsymbol{\zeta}_k, \Omega_k)$ $(1\leq k \leq N)$ from the posterior distribution
  \begin{equation*}
  f_{\mathrm{Normal}}(\boldsymbol{\zeta}_k \, \vert \, \boldsymbol{m}, \frac{1}{\lambda} \Omega_k)\times f_{\mathrm{IW}}(\Omega_k \, \vert \, \nu, \Psi) \times \prod\limits_{\{i:\kappa_i = k\}} f_{\mathrm{Normal}}(\boldsymbol{z}_i \, \vert \, \boldsymbol{\zeta}_k, \Omega_k)
  \end{equation*}
  which essentially is (\ref{eq:param}-\ref{eq:param1}), if $k \in \boldsymbol{\kappa}^*$, and sample a new value for each $(\boldsymbol{\zeta}_k, \Omega_k)$ from the prior distribution $f_{\mathrm{Normal}}(\boldsymbol{\zeta}_{k}\,\vert\,\boldsymbol{m},\frac{1}{\lambda}\Omega_{k}) \times f_{\mathrm{IW}}(\Omega_{k}\,\vert\,\nu,\Psi)$ otherwise.
  \item Sample a vector of new values for $\{\omega_k\}_{k=1}^N$ from
  \begin{align*}
  f(\boldsymbol{\omega} \, \vert \, \{\kappa_i\}_{i=1}^n, \alpha) & \, \propto \, f_{\mathrm{GD}}(\boldsymbol{\omega} \, \vert \, 1, \alpha, \cdots, 1, \alpha) \times \prod\limits_{i=1}^n f_{\mathrm{Discrete}}(\kappa_i \, \vert \, \boldsymbol{\omega}) \\
  & \, \propto \, f_{\mathrm{GD}}(\boldsymbol{\omega} \, \vert \, a_1, b_1, \cdots, a_{N-1}, b_{N-1})
  \end{align*}
  using two steps:
  \begin{enumerate}
    \item sample a value for $V_k$ from $\mathrm{Beta}(a_k, b_k)$ for $1 \leq k \leq N-1$ and set $V_N = 1$,
    \item compute $\omega_1 = V_1$ and $\omega_k = (1-V_1)\cdots(1-V_{k-1})V_k$ for $2 \leq k \leq N$,
  \end{enumerate}
  where $a_k = n_k + 1$, $b_k = N_k + \alpha$, $N_k = n_{k+1} + \cdots + n_N$ for $1 \leq k \leq N-1$ and $n_k = \vert \{i: \kappa_i = k \}\vert$ for $1 \leq k \leq N$.
  \item Sample a new value for each $\kappa_i (1\leq i \leq n)$ from a discrete distribution
  \begin{align*}
  P(\kappa_i = k \, \vert \, \boldsymbol{\omega}, \{(\boldsymbol{\zeta}_k, \Omega_k)\}_{k=1}^{N}, \{\boldsymbol{z}_i\}_{i=1}^n) & \, \propto \,  P(\kappa_i = k \, \vert \, \boldsymbol{\omega}) \times f_{\mathrm{Normal}}(\boldsymbol{z}_i \, \vert \, \boldsymbol{\zeta}_k, \Omega_k) \\
  & \, = \, \omega_k f_{\mathrm{Normal}}(\boldsymbol{z}_i \, \vert \, \boldsymbol{\zeta}_k, \Omega_k) 
  \end{align*}
  where $k = 1, \cdots, N$. In practice, Gumbei-max trick (\cite{maddison2014sampling}) is used to help sample the discrete distribution in log-space and avoid potentially floating point inaccuracy.
  \item Sample a new value for $\alpha$ from
  \begin{align*}
  f(\alpha \, \vert \, \boldsymbol{\omega}, a_0, b_0) & \, \propto \, f_{\mathrm{Gamma}}(\alpha \, \vert \, a_0, b_0) \times f_{\mathrm{GD}}(\boldsymbol{\omega} \, \vert \, 1, \alpha, \cdots, 1, \alpha) \\
  & \, \propto \, f_{\mathrm{Gamma}}(\alpha \, \vert \, a_0^*, b_0^*),
  \end{align*}
  where $a_0^* = a_0 + N - 1$ and $b_0^* = b_0 - \log(\omega_N) = b_0 - \sum\limits_{k=1}^{N-1} \log (1-V_k)$.
  \item Sample a new value for $\boldsymbol{m}$ from (\ref{eq:m}) with $\{\boldsymbol{\zeta}_k, \Omega_k\}_{k \in \{\kappa_i \}_{i=1}^n }$ replaced by $\{\boldsymbol{\zeta}_k, \Omega_k\}_{k=1}^{N}$.
  \item Sample a new value for $\lambda$ from (\ref{eq:lambda}) with $\{\boldsymbol{\zeta}_k, \Omega_k\}_{k \in \{\kappa_i \}_{i=1}^n }$ replaced by $\{\boldsymbol{\zeta}_k, \Omega_k\}_{k=1}^{N}$ and $K^*$ replaced by $N$.
  \item Sample a new value for $\Psi$ from (\ref{eq:psi}) with $\{\boldsymbol{\zeta}_k, \Omega_k\}_{k \in \{\kappa_i \}_{i=1}^n }$ replaced by $\{\boldsymbol{\zeta}_k, \Omega_k\}_{k=1}^{N}$ and $K^*$ replaced by $N$.
\end{enumerate}
We can see that steps 5-7 are similar to steps 2-4 in Appendix~\ref{app:neal} except that steps 2-4 in Appendix~\ref{app:neal} update hyper-parameters of $G_0$ using $(\boldsymbol{\zeta}_k, \Omega_k)$'s associated with at least one observation while steps 5-7 update hyper-parameters of $G_0$ using all the $(\boldsymbol{\zeta}_k, \Omega_k)$'s. Since some of $\{\boldsymbol{\zeta}_k, \Omega_k\}_{k=1}^{N}$ are not related to any observations, the Markov chain of hyper-parameters will mix slowly if the blocked Gibbs sampler is used.


\section{Density Estimation Based on Pólya Urn Gibbs Sampler} \label{app:estimation}

In this section, we describe the density estimation approach of \cite{muller1996bayesian} using Algorithm 8 of \cite{neal2000markov}.

The predictive density $f(\boldsymbol{z}_{n+1} \, \vert \, \{\boldsymbol{z}_i\}_{i=1}^n)$ can be expressed as follows
\begin{align} \label{eq:d1}
f(\boldsymbol{z}_{n+1} \, \vert \, \{\boldsymbol{z}_i\}_{i=1}^n) \, = \, \int \int & f(\boldsymbol{z}_{n+1} \, \vert \, \boldsymbol{\mu}_{n+1}, \Sigma_{n+1}) f(\boldsymbol{\mu}_{n+1}, \Sigma_{n+1} \, \vert \, \{\boldsymbol{\zeta}_k, \Omega_k\}_{k \in \{\kappa_i\}_{i=1}^n}) d(\boldsymbol{\mu}_{n+1}, \Sigma_{n+1}) \nonumber \\
& d F(\{\boldsymbol{\zeta}_k, \Omega_k\}_{k \in \{\kappa_i\}_{i=1}^n} \, \vert \, \{\boldsymbol{z}_i\}_{i=1}^n),
\end{align}
where $f(\boldsymbol{z}_{n+1} \, \vert \, \boldsymbol{\mu}_{n+1}, \Sigma_{n+1})$ is the normal likelihood, $F(\{\boldsymbol{\zeta}_k, \Omega_k\}_{k \in \{\kappa_i\}_{i=1}^n} \, \vert \, \{\boldsymbol{z}_i\}_{i=1}^n)$ is the posterior distribution, and $f(\boldsymbol{\mu}_{n+1}, \Sigma_{n+1} \, \vert \, \{\boldsymbol{\zeta}_k, \Omega_k\}_{k \in \{\kappa_i\}_{i=1}^n})$ can be obtained from (\ref{eq:6}) with $G$ integrated out and can be represented using the following Pólya Urn scheme:
\begin{align*}
f(\boldsymbol{\mu}_{n+1}, \Sigma_{n+1} \, \vert \, \{\boldsymbol{\zeta}_k, \Omega_k\}_{k \in \{\kappa_i\}_{i=1}^n})  \, = \, & \frac{\alpha}{\alpha + n}G_0(\boldsymbol{\mu}_{n+1}, \Sigma_{n+1})\\
& +  \sum\limits_{k \in \{\kappa_i\}_{i=1}^n} \frac{n_k}{\alpha + n}\delta_{(\boldsymbol{\zeta}_k, \Omega_k)} (\boldsymbol{\mu}_{n+1}, \Sigma_{n+1}).
\end{align*}
Thus, with each posterior sample $\{ \{\kappa_i^{\{l\}}\}_{i=1}^n, \{\boldsymbol{\zeta}_k^{\{l\}}, \Omega_k^{\{l\}}\}_{k \in \{\kappa_i^{\{l\}}\}_{i=1}^n} \}$ from the Pólya urn Gibbs sampler, the inside integral of (\ref{eq:d1}) can be estimated by
\begin{equation} \label{eq:d2}
\frac{\alpha}{\alpha + n} f(\boldsymbol{z}_{n+1} \, \vert \, \boldsymbol{\zeta}^{*}, \Omega^{*}) + \sum\limits_{k \in \{\kappa_i^{\{l\}}\}_{i=1}^n} \frac{n_k}{\alpha + n} f(\boldsymbol{z}_{n+1} \, \vert \, \boldsymbol{\zeta}_{k}^{\{l\}}, \Omega_{k}^{\{l\}}),
\end{equation}
where $(\boldsymbol{\zeta}^{*}, \Omega^{*})$ is sampled from the base distribution $G_0$. The predictive density (\ref{eq:d1}) can be estimated by taking the average of (\ref{eq:d2}) across all the MCMC samples. The derivation above assumes that $\alpha$ and hyper-parameters of $G_0$ are fixed. If the hierarchical model (\ref{eq:8}) is assumed, the estimator in (\ref{eq:d2}) can be adjusted by using the latest update of $\alpha$ and the hyper-parameters.

When the Pólya urn Gibbs sampler is used, we use the \fct{DPcdensity} function in \pkg{DPpackage} to estimate the predictive conditional density $f(Y_{n+1} \, \vert \, \boldsymbol{x}_{n+1}, \{\boldsymbol{z}_i\}_{i=1}^n)$ using an extension of the approach of \cite{muller1996bayesian}. They first fit a DPM of multivariate normals model on the joint data $\{\boldsymbol{z}_i\}_{i=1}^n$ and get a sequence of posterior samples. Then a weight dependent mixture model is induced from the joint DPM model. Based on the stick-breaking representation of DP in (\ref{eq:11}), the weight dependent mixture model has the following form
\begin{align} \label{eq:d3}
f(Y_{n+1} \, \vert \, \boldsymbol{x}_{n+1}, \{\boldsymbol{z}_i\}_{i=1}^n) \, = \, \int & \left( \sum\limits_{j=1}^{\infty} \omega_j(\boldsymbol{x}_{n+1}) f_{\mathrm{Normal}} (Y_{n+1} \, \vert \, \beta_{0j} + \boldsymbol{x}_{n+1}^{\top} \boldsymbol{\beta}_j, \sigma_j^2) \right) \nonumber \\
& dF(\mathcal{P}_{\infty} \, \vert \, \{\boldsymbol{z}_i\}_{i=1}^n),
\end{align}
where $\omega_j(\boldsymbol{x}_{n+1}), \beta_{0j}, \boldsymbol{\beta}_j$ and $\sigma_j^2$ are functions of the components of $\mathcal{P}_{\infty}$, i.e., $\{\omega_j\}_{j=1}^{\infty}, \{\kappa_i \}_{i=1}^n$ and $\{\boldsymbol{\zeta}_k, \Omega_k\}_{k \in \{\kappa_i\}_{i=1}^n}$. Finally, the $\epsilon$-DP approximation proposed by \cite{muliere1998approximating} with $\epsilon = 0.01$ is used for the inference of (\ref{eq:d3}). The detailed procedure of estimating the summation in (\ref{eq:d3}) is summarized in Algorithm \ref{alg:3}.

\begin{algorithm}
\caption{Weight Dependent DPMM for Conditional Predictive Density}\label{alg:3}
\begin{algorithmic}[1]

\Procedure{Estimate the sum in (\ref{eq:d3}) given a posterior sample $\{\kappa_i, \boldsymbol{\zeta}_{\kappa_i}, \Omega_{\kappa_i} \}_{i=1}^n$ from the joint DPM model with $\epsilon=0.01$}{}
    \State Label $\{\kappa_1, \cdots, \kappa_n\}$ with $\{1, \cdots, K^*\}$
    \For{$k \in \{1, \cdots, K^* \}$}
        \State Calculate $\tilde{\beta}_{0k}, \, \tilde{\boldsymbol{\beta}}_k$ and $\tilde{\sigma}_k^2$ as follows:
        \begin{align} \label{eq:d4}
        & \tilde{\boldsymbol{\beta}}_k \, = \, \Omega_{12k} \Omega_{22k}^{-1}, \quad \tilde{\beta}_{0k} \, = \, \zeta_{1k} - \tilde{\boldsymbol{\beta}}_k \boldsymbol{\zeta}_{2k}, \quad \tilde{\sigma}_k^2  \, = \, \Omega_{11k} - \tilde{\boldsymbol{\beta}}_k \Omega_{21k}, \\
        & \boldsymbol{\zeta}_{k} \,=\,\left(\begin{array}{c}
        \zeta_{1k}\\
        \boldsymbol{\zeta}_{2jk}
        \end{array}\right) \quad \text{and} \quad \Omega_{k}\,=\,\left(\begin{array}{cc}
        \Omega_{11k} & \Omega_{12k}\\
        \Omega_{21k} & \Omega_{22k}
        \end{array}\right)  \nonumber
        \end{align}
        \State Calculate $f_{\mathrm{Normal}}(\boldsymbol{x}_{n+1} \, \vert \, \boldsymbol{\zeta}_{2k}, \Omega_{22k})$ and $f_{\mathrm{Normal}} (Y_{n+1} \, \vert \, \tilde{\beta}_{0k} + \boldsymbol{x}_{n+1}^{\top} \tilde{\boldsymbol{\beta}}_k, \tilde{\sigma}_k^2)$ \label{step:5}
    \EndFor
    \State Set $\omega_0 = 0$ and $j = J = 0$
    \While{$1-\sum\limits_{j=0}^J \omega_j > \epsilon$}
        \State Update $j = j + 1$ and $J = J + 1$
        \State Sample $V_j$ from $\mathrm{Beta}(1, \alpha+n)$, and compute $\omega_j = V_j\prod\limits_{s=0}^{j-1} (1-V_{s-1})$
        \State Sample $k_j$ from $\{1, \cdots, K^*+1 \}$ with probabilities $\{\frac{n_1}{\alpha+n},\cdots, \frac{n_{K^*}}{\alpha+n}, \frac{\alpha}{\alpha+n}\}$
        \If{$k_j \leq K^*$}
            \State Set $\beta_{0j} = \tilde{\beta}_{0k_j}, \, \boldsymbol{\beta}_j = \tilde{\boldsymbol{\beta}}_{k_j}$ and $\sigma_j^2 = \tilde{\sigma}_{k_j}^2$
            \State Use $f_{\mathrm{Normal}}(\boldsymbol{x}_{n+1} \, \vert \, \boldsymbol{\zeta}_{2k_j}, \Omega_{22k_j})$ in step \ref{step:5} to calculate $\tilde{\omega}_j(\boldsymbol{x}_{n+1})$ as follows:
            \begin{equation} \label{eq:d5}
            \tilde{\omega}_j(\boldsymbol{x}_{n+1}) \, =  \, \omega_j f_{\mathrm{Normal}}(\boldsymbol{x}_{n+1} \, \vert \, \boldsymbol{\zeta}_{2k_j}, \Omega_{22k_j})
            \end{equation}
        \Else
            \State Sample $(\boldsymbol{\zeta}^*, \Omega^*)$ from the base distribution $G_0$
            \State Calculate $\beta_{0j}, \, \boldsymbol{\beta}_j, \, \sigma_j^2$ and $\tilde{\omega}_j(\boldsymbol{x}_{n+1})$ using (\ref{eq:d4}) and (\ref{eq:d5}), respectively, with $(\boldsymbol{\zeta}_k, \Omega_k)$ replaced by $(\boldsymbol{\zeta}^*, \Omega^*)$
        \EndIf
    \EndWhile
    \State The sum in (\ref{eq:d3}) is estimated by
    \begin{equation} \label{eq:d6}
    \frac{\sum\limits_{j=1}^J \tilde{\omega}_{j}(\boldsymbol{x}_{n+1}) f_{\mathrm{Normal}} (Y_{n+1} \, \vert \, \beta_{0j} + \boldsymbol{x}_{n+1}^{\top} \boldsymbol{\beta}_j, \sigma_j^2)}{\sum\limits_{j=1}^J \tilde{\omega}_{j}(\boldsymbol{x}_{n+1})}
    \end{equation}
    and notice that for those $j$ such that $k_j \leq K^*$, the conditional density $f_{\mathrm{Normal}} (Y_{n+1} \, \vert \, \beta_{0j} + \boldsymbol{x}_{n+1}^{\top} \boldsymbol{\beta}_j, \sigma_j^2)$ is computed in step \ref{step:5}
\EndProcedure

\end{algorithmic}
\end{algorithm}
The estimation of a predictive conditional density in (\ref{eq:d3}) based on the Pólya urn sampler has similar construction of that in (\ref{eq:17}) based on the blocked Gibbs sampler. However, it is important to note that the one based on the Pólya urn sampler is much more complicated and computationally intensive, because it does not have the direct access to the mixing weights. To estimate a predictive conditional density, Algorithm \ref{alg:3} has to repeatedly generate stick-breaking weights to satisfy the conditions of $\epsilon$-DP approximation, after each MCMC iteration of the DPM model. This typically involves thousands of stick-breaking weights, which implies that it takes thousands of basic operations to evaluate the density for only one data point.


\section{Marginal Partition Posterior and Likelihood for DPM Models} \label{app:diag}

The likelihood functions given the data $\{\boldsymbol{z}_i \}_{i=1}^n$ under the DPM of multivariate normals model fitted by the Pólya urn Gibbs sampler (Algorithm~\ref{alg:1}) and the blocked Gibbs sampler (Algorithm~\ref{alg:2}) have the same form as follows,
\begin{align*}
L\left(\{\kappa_i\}_{i=1}^n, \{\boldsymbol{\zeta}_k, \Omega_k  \}_{k=1}^{N^*} \, \vert \, \{\boldsymbol{z}_i \}_{i=1}^n \right) & \, = \, f\left(\{\boldsymbol{z}_i \}_{i=1}^n  \, \vert \, \{\kappa_i\}_{i=1}^n, \{\boldsymbol{\zeta}_k, \Omega_k \}_{k=1}^{N^*} \right)   \\
& \, = \,  \prod\limits_{j=1}^{K^*} \left[ \prod\limits_{\{ i: \kappa_i = \kappa_j^* \}} f_{\mathrm{Normal}}(\boldsymbol{z}_i \, \vert \, \boldsymbol{\zeta}_{\kappa_j^*}, \Omega_{\kappa_j^*}) \right],
\end{align*}
where $N^*$ is the pre-specified number of clusters $N$ if the blocked Gibbs sampler is used and is the number of non-empty clusters $K^*$ if the Pólya urn Gibbs sampler is used, and $\boldsymbol{\kappa}^* = \{\kappa_j^*\}_{j=1}^{K^*}$ are the unique values in $\{\kappa_i\}_{i=1}^n$.

The marginal partition posterior is only available for the blocked Gibb sampling. Under Model~(\ref{eq:14}), the marginal partition posterior is defined as follows
\begin{equation} \label{eq:e1}
f(\kappa_1, \cdots, \kappa_n \, \vert \, \{\boldsymbol{z}_i \}_{i=1}^n ) \propto f(\{\boldsymbol{z}_i \}_{i=1}^n  \, \vert \, \kappa_1, \cdots, \kappa_n) \times f(\kappa_1, \cdots, \kappa_n).
\end{equation}

The first term on the right hand side is the marginal likelihood with $\boldsymbol{\zeta}_k$'s and $\Omega_k$'s integrated out, and can be calculated as follows
\begin{align*}
f(\{\boldsymbol{z}_i \}_{i=1}^n  \, \vert \, \kappa_1, \cdots, \kappa_n) & \,  =  \, \int \prod\limits_{j=1}^{K^*} \left[ f_{\mathrm{NIW}}(\boldsymbol{\zeta}_{\kappa_j^*}, \Omega_{\kappa_j^*} \, \vert \, \boldsymbol{m}, \lambda, \nu, \Psi)  \prod\limits_{\{ i: \kappa_i = \kappa_j^* \}} f_{\mathrm{Normal}}(\boldsymbol{z}_i \, \vert \, \boldsymbol{\zeta}_{\kappa_j^*}, \Omega_{\kappa_j^*}) \right] \\
& \quad  \quad d \{\boldsymbol{\zeta}_{\kappa_j^*}, \Omega_{\kappa_j^*} \}_{j=1}^{K^*} \times \int \prod\limits_{k \notin \boldsymbol{\kappa}^*} f_{\mathrm{NIW}}(\boldsymbol{\zeta}_k, \Omega_k \, \vert \, \boldsymbol{m}, \lambda, \nu, \Psi) d \{\boldsymbol{\zeta}_k, \Omega_k \}_{k \notin \boldsymbol{\kappa}^*} \\
& \, = \, \frac{\vert \Psi \vert^{\frac{K^* \nu}{2}}}{\pi^{\frac{nd}{2}}\Gamma^m_d(\frac{\nu}{2})} \prod\limits_{j=1}^{K^*} \frac{\Gamma_d(\frac{\nu + n_j}{2}) \left( \frac{\lambda}{\lambda + n_j} \right)^{\frac{d}{2}} }{\vert \Psi^* \vert^{\frac{\nu + n_j}{2}}},
\end{align*}
where $n_j$ and $\Psi^*$ are defined in (\ref{eq:param1}) and $\Gamma_d(\cdot)$ is a multivariate gamma function with dimension $d$. Note that in the first equality above, the second integral is $1$, and the first integral is a product of $K^*$ marginal likelihoods of $\{\boldsymbol{z}_i \}_{i=1}^n$ given $K^*$ conjugate normal-inverse-Wishart priors, respectively.

The second term on the right hand side of (\ref{eq:e1}) is the marginal prior distribution of $\{\kappa_i\}_{i=1}^n$, and it can be computed as follows
$$f(\kappa_1, \cdots, \kappa_n) \, = \, \int \int f(\kappa_1, \cdots, \kappa_n \, \vert \, \boldsymbol{\omega}) f(\boldsymbol{\omega} \, \vert \, \alpha) f(\alpha \, \vert \,  a_0, b_0) d \boldsymbol{\omega} d \alpha  .$$
For simplicity, we fix $\alpha$, for example, at its posterior mean $\alpha^*$, so the density of interest can be calculated as follows
\begin{align*}
f(\kappa_1, \cdots, \kappa_n) & \, = \, \int f(\kappa_1, \cdots, \kappa_n \, \vert \, \boldsymbol{\omega}) f(\boldsymbol{\omega} \, \vert \, \alpha^*) d \boldsymbol{\omega} \\
& \, = \, \int \left( \prod\limits_{j=1}^{K^*} \omega_{\kappa_j^*}^{n_j} \right) \times f_{\mathrm{GD}}(\omega_1,\cdots,\omega_N  \, \vert \, 1, \alpha^*, \cdots, 1, \alpha^*) d \boldsymbol{\omega} \\
& \, = \, \frac{\alpha^{*N-1} \Gamma(b_{N-1}) \prod\limits_{k=1}^{N-1}  \Gamma(a_k) }{\Gamma(n + \alpha^* + 1) \prod\limits_{k=1}^{N-2} b_k},
\end{align*}
where $a_k$ and $b_k$ are defined in the step 2 of Appendix~\ref{app:truncated}. The last equality of the equation above is from the conjugacy of the generalized Dirichlet distribution and the multinomial distribution.


\section{More Simulation Results} \label{app:simulation}

%
\begin{figure}[h!]
\centering
\includegraphics{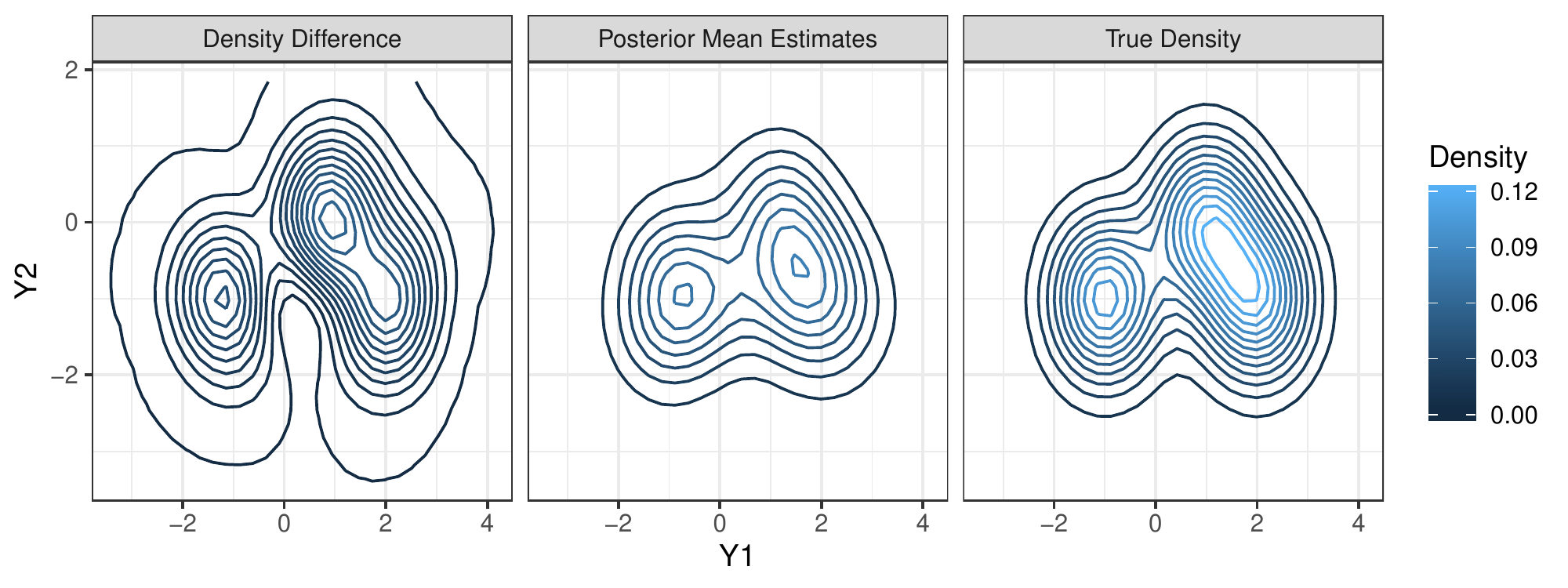}
\caption{\label{fig:jt2}From left to right, subfigures are contours of the difference between the estimated density and the true density, the estimated density, and the true density, for the DPM of multivariate normals model using the Pólya urn Gibbs sampler for the example in (\ref{exp:2}).}
\end{figure}
%

%
\begin{figure}[h!]
\centering
\includegraphics{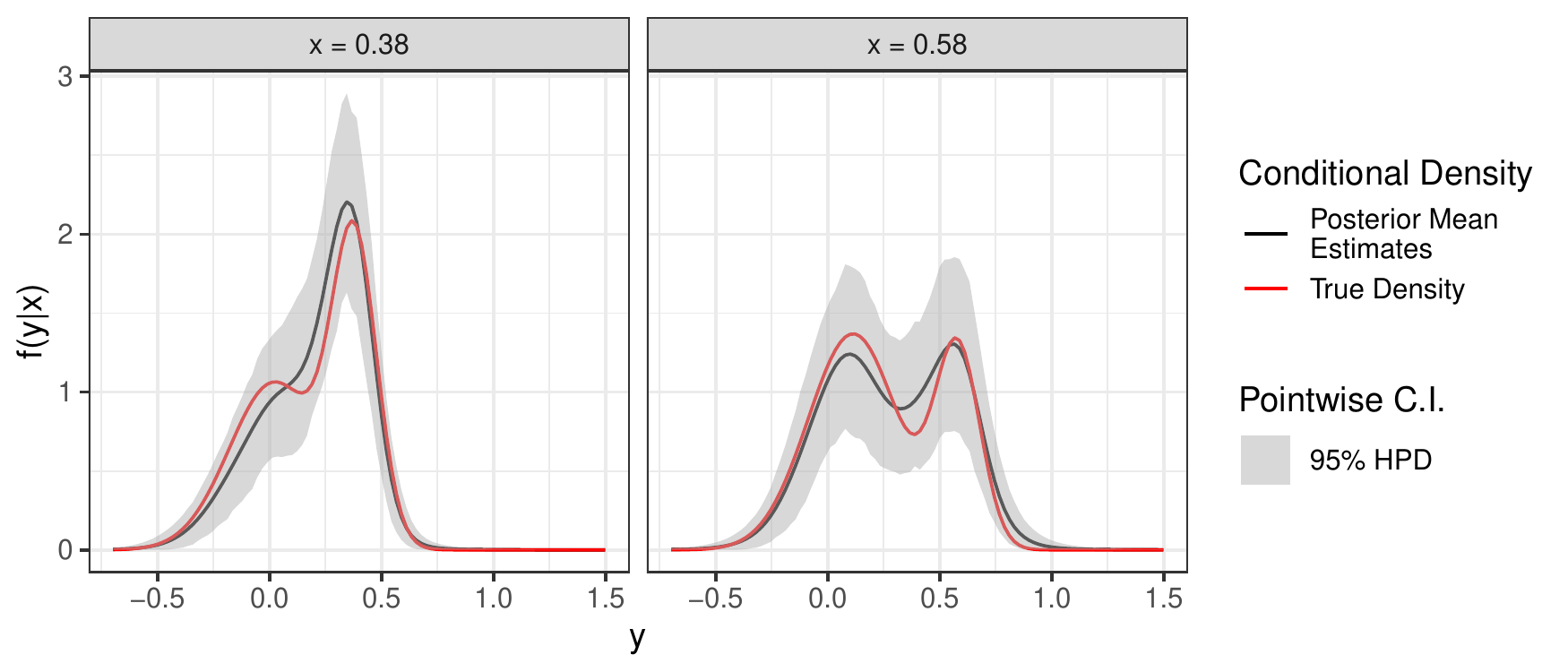}
\caption{\label{fig:cd4}Estimated conditional density for the DPM of multivariate normals model using the Pólya urn Gibbs sampler for the example in (\ref{exp:3}).}
\end{figure}
\begin{figure}[h!]
\centering
\includegraphics{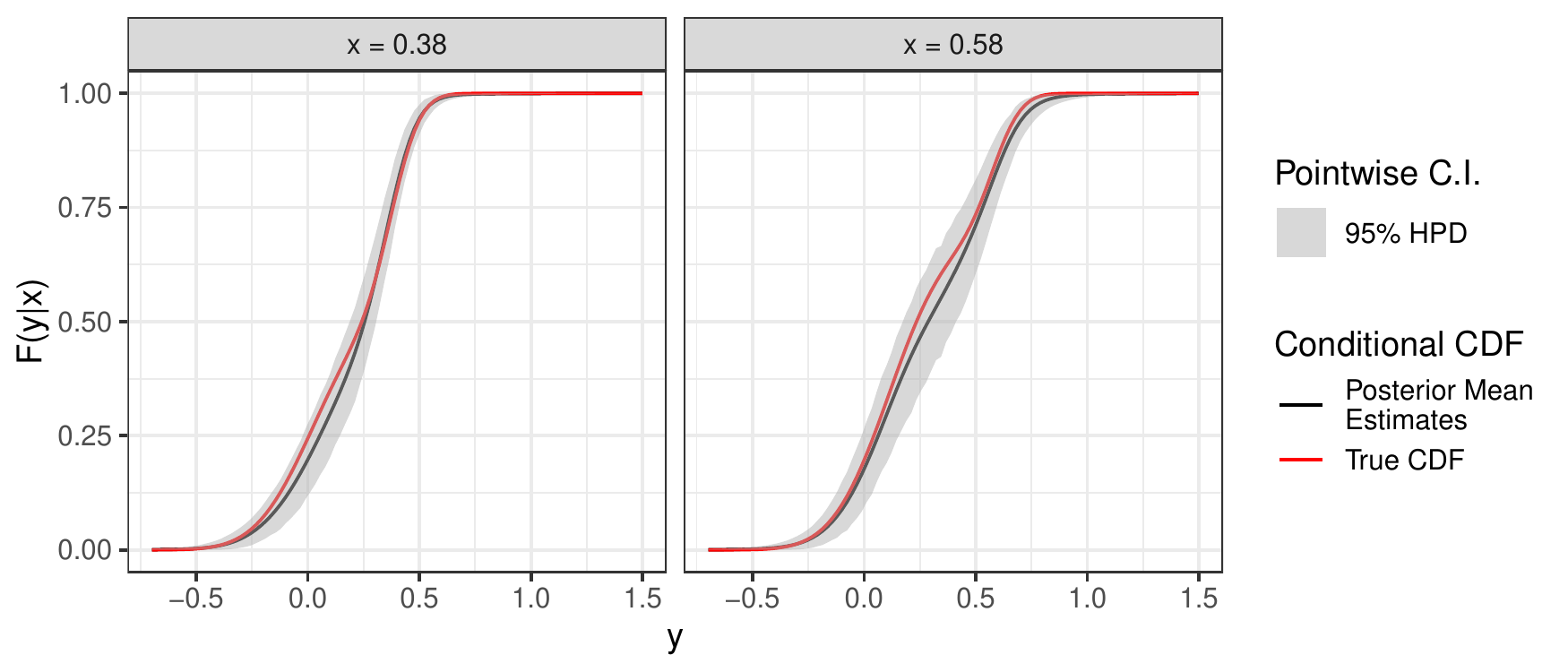}
\caption{\label{fig:cd5}Estimated conditional CDF for the DPM of multivariate normals model using the Pólya urn Gibbs sampler for the example in (\ref{exp:3}).}
\end{figure}
\begin{figure}[h!]
\centering
\includegraphics{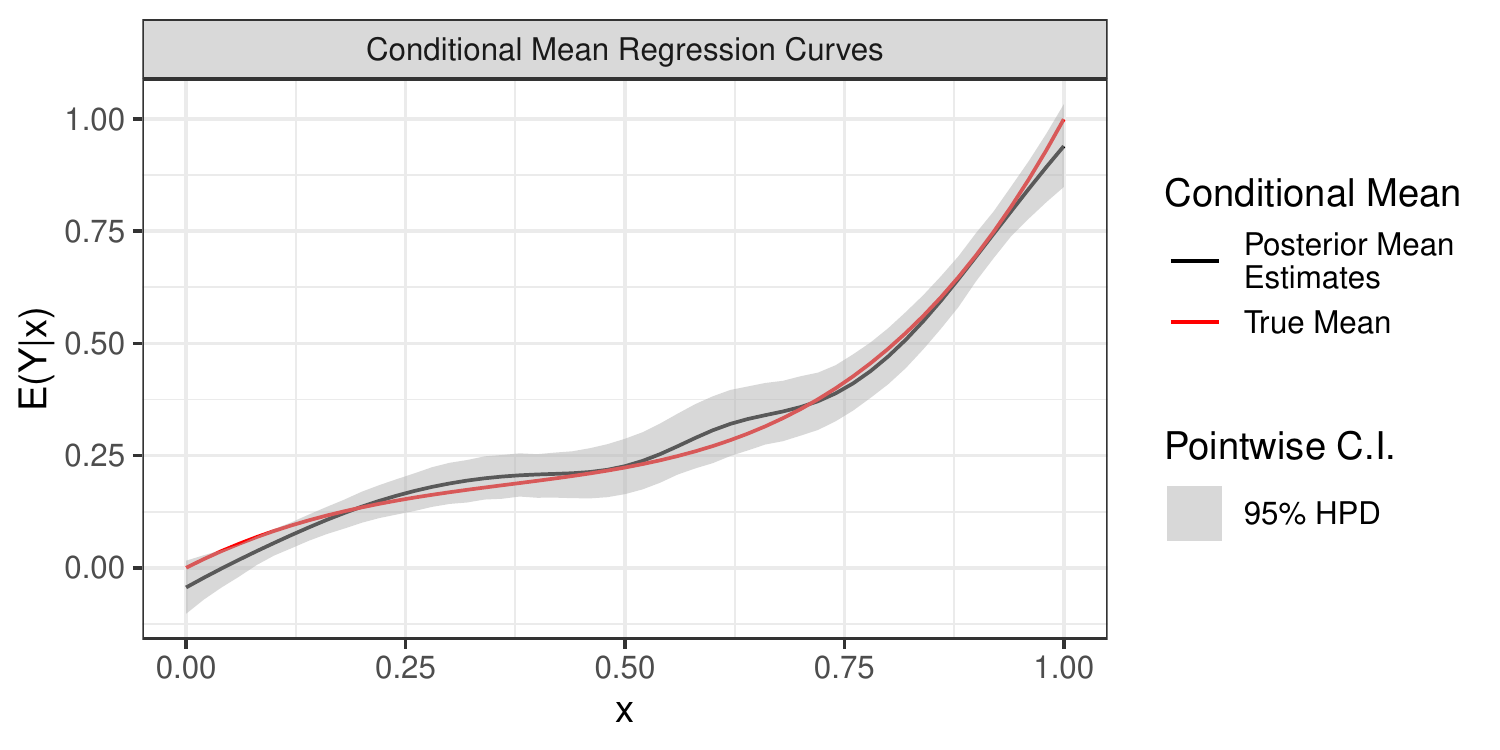}
\caption{\label{fig:cd6}Estimated conditional mean for the DPM of multivariate normals model using the Pólya urn Gibbs sampler for the example in (\ref{exp:3}).}
\end{figure}

\end{appendix}

\end{document}